%% file: cuore0-background_v1.tex
\pdfoutput=1
\pdfoutput=1
\pdfoutput=1
\pdfoutput=1
\pdfoutput=1
\RequirePackage{fix-cm}
\documentclass[twocolumn,epjc3-upd,fleqn]{svjour3}
\smartqed  
\RequirePackage{graphicx}
\RequirePackage{subfigure}
\usepackage{rotating}
\usepackage{amsmath}
\usepackage{mathtools}
\RequirePackage[colorlinks,citecolor=blue,urlcolor=blue,linkcolor=blue]{hyperref}
\usepackage{doi}
\usepackage[]{units}
\usepackage{xspace}
\usepackage[switch]{lineno} 
\journalname{Eur. Phys. J. C}
\input{definitions.tex}
\begin{document}
\title{Measurement of the Two-Neutrino Double Beta Decay Half-life of \tect with the CUORE-0 Experiment}
\input{authorlist_27May2016.tex}
\date{Received: date / Accepted: date}
\twocolumn
\maketitle
\begin{abstract}
We report on the measurement of the two-neutrino double beta decay half-life of \tect with the \qz detector. From an exposure of \unit[33.4]{kg$\cdot$y} of \teod, the half-life is determined to be \Tdn~ = [8.2 $\pm$ 0.2 (stat.) $\pm$ 0.6  (syst.)] $\times$ 10$^{20}$~y. 
This result is obtained after a detailed reconstruction of the sources responsible for the \qz\ counting rate, with a specific study of those contributing to the \tect neutrinoless double beta decay region of interest.

\end{abstract}

\section{Introduction}
\label{sec:intro}
\input{introduction.tex}

\section{Experiment}
\label{sec:exp}
\input{exp.tex}

\section{Background Sources}
\label{sec:sources}
\input{bkg_sources}

\section{Monte Carlo}
\label{sec:MC}
\input{montecarlo.tex}

\section{Reconstruction of the Calibration Source} 
\label{sec:calibration}
\input{calibration.tex}

\section{Construction of Background Model}
\label{sec:bkg_model}
\input{bkg_model.tex}

\section{Bayesian Fit Construction}
\label{sec:jags}
\input{JAGS_pl.tex}

\section{Reference Fit and Systematics}
\label{sec:systematics}
\input{systematics.tex}

\input{results.tex}

\section{Conclusion}
\input{conclusion.tex}

\section{Acknowledgments}
The CUORE Collaboration thanks the directors and staff of the
Laboratori Nazionali del Gran Sasso and the technical staff of our
laboratories. This work was supported by the Istituto Nazionale di
Fisica Nucleare (INFN); the National Science
Foundation under Grant Nos. NSF-PHY-0605119, NSF-PHY-0500337,
NSF-PHY-0855314, NSF-PHY-0902171, NSF-PHY-0969852, NSF-PHY-1307204, NSF-PHY-1314881, NSF-PHY-1401832, and NSF-PHY-1404205; the Alfred
P. Sloan Foundation; the University of Wisconsin Foundation; and Yale
University. This material is also based upon work supported  
by the US Department of Energy (DOE) Office of Science under Contract Nos. DE-AC02-05CH11231,
DE-AC52-07NA27344, and DE-SC0012654; and by the DOE Office of Science, Office of Nuclear Physics under Contract Nos. DE-FG02-08ER41551 and DE-FG03-00ER41138.
This research used resources of the National Energy Research Scientific Computing Center (NERSC).

\bibliographystyle{spphys}       
\bibliography{bib}   

\end{document}

%% file: definitions.tex
\def\qz{\mbox{CUORE-0}\xspace}
\def\qino{\mbox{Cuoricino}\xspace}

\def\mcuorez{\emph{MCuoreZ}\xspace} 
\def\pbext{\emph{ExtPb}\xspace} 
\def\cryoext{\emph{CryoExt}\xspace} 
\def\cryoint{\emph{CryoInt}\xspace} 
\def\pbint{\emph{IntPb}\xspace} 
\def\holder{\emph{Holder}\xspace}
\def\smallpart{\emph{Small Parts}\xspace}
\def\crystal{\emph{Crystals}\xspace}

\def\mspec{$\mathcal{M}$\xspace}
\def\mspecone{$\mathcal{M}_1$\xspace}
\def\mspectwo{$\mathcal{M}_2$\xspace}

\def\summspectwo{$\Sigma_2$\xspace}

\def\u{$^{238}$U\xspace}
\def\ths{$^{232}$Th\xspace}
\def\thdto{$^{230}$Th\xspace}
\def\acddo{$^{228}$Ac\xspace}
\def\raddq{$^{224}$Ra\xspace}
\def\rndd{$^{222}$Rn\xspace}

\def\bidq{$^{214}$Bi\xspace}
\def\pbduq{$^{214}$Pb\xspace}
\def\podud{$^{212}$Po\xspace}
\def\bidud{$^{212}$Bi\xspace}
\def\pbdud{$^{212}$Pb\xspace}
\def\podd{$^{210}$Po\xspace}
\def\pbdd{$^{210}$Pb\xspace}
\def\tld{$^{208}$Tl\xspace}
\def\bidzs{$^{207}$Bi\xspace}
\def\pt{$^{190}$Pt\xspace}

\def\csuts{$^{137}$Cs\xspace}

\def\tect{$^{130}$Te\xspace}

\def\teudcm{$^{125m}$Te\xspace}
\def\sbudc{$^{125}$Sb\xspace}

\def\aguzom{$^{108m}$Ag\xspace}

\def\cosz{$^{60}$Co\xspace}
\def\cocs{$^{57}$Co\xspace}
\def\mncq{$^{54}$Mn\xspace}

\def\kq{$^{40}$K\xspace}

\def\teod{TeO$_2$\xspace}

\def\gm{$\gamma$\xspace}

\def\alph{$\alpha$\xspace}
\def\bmin{$\beta^-$\xspace}

\def\bbd{$2\nu\beta\beta$\xspace}
\def\bbz{$0\nu\beta\beta$\xspace}

\def\Tdn{$T_{1/2}^{2\nu}$}

\def\vita-dim{$\tau_{1/2}$}

\def\ca{$\sim$}

\def\pom{$\pm$ }

\def\be{\begin{equation}}
\def\ee{\end{equation}}

\def\ums{$\mu$m}

%% file: authorlist_27May2016.tex
\author{C.~Alduino\thanksref{USC} 
\and
K.~Alfonso\thanksref{UCLA} 
\and
D.~R.~Artusa\thanksref{USC,LNGS} 
\and
F.~T.~Avignone~III\thanksref{USC} 
\and
O.~Azzolini\thanksref{INFNLegnaro} 
\and
T.~I.~Banks\thanksref{BerkeleyPhys,LBNLNucSci} 
\and
G.~Bari\thanksref{INFNBologna} 
\and
J.~W.~Beeman\thanksref{LBNLMatSci} 
\and
F.~Bellini\thanksref{Roma,INFNRoma} 
\and
A.~Bersani\thanksref{INFNGenova}
% ...
} 
\institute{Department of Physics and Astronomy, University of South Carolina, Columbia, SC 29208 - USA\label{USC} 
\and
Department of Physics and Astronomy, University of California, Los Angeles, CA 90095 - USA\label{UCLA} 
\and
INFN - Laboratori Nazionali del Gran Sasso, Assergi (L'Aquila) I-67010 - Italy\label{LNGS} 
\and
INFN - Laboratori Nazionali di Legnaro, Legnaro (Padova) I-35020 - Italy\label{INFNLegnaro} 
\and
Department of Physics, University of California, Berkeley, CA 94720 - USA\label{BerkeleyPhys} 
\and
Nuclear Science Division, Lawrence Berkeley National Laboratory, Berkeley, CA 94720 - USA\label{LBNLNucSci} 
\and
INFN - Sezione di Bologna, Bologna I-40127 - Italy\label{INFNBologna} 
\and
Materials Science Division, Lawrence Berkeley National Laboratory, Berkeley, CA 94720 - USA\label{LBNLMatSci} 
\and
Dipartimento di Fisica, Sapienza Universit\`{a} di Roma, Roma I-00185 - Italy\label{Roma} 
\and
INFN - Sezione di Roma, Roma I-00185 - Italy\label{INFNRoma} 
\and
INFN - Sezione di Genova, Genova I-16146 - Italy\label{INFNGenova} 
% ...
}

\thankstext{e1}{E-mail: cuore-spokesperson@lngs.infn.it}
\thankstext{n1}{Deceased}

%% file: introduction.tex
Double-beta decay is a rare nuclear process in which two nucleons simultaneously decay and emit two electrons. The allowed Standard Model version of this process emits two (anti-)neutrinos and is called two-neutrino double-beta decay (\bbd). This decay is interesting in its own right as the slowest process ever directly observed~\cite{Exo,Kz}. Moreover it may represent an important source of background for the neutrinoless double-beta decay (\bbz), i.e. a related process with no neutrino emission~\cite{Furry}. 
\bbz manifestly violates lepton number and therefore its discovery would point to new physics beyond the Standard Model. Experiments searching for \bbz have made great leaps forward in sensitivity by using a variety of techniques and isotopes~\cite{Exo,Kz,Gerda,Us}.

The Cryogenic Underground Observatory for Rare Events (CUORE)~\cite{Us1,Us2} is the latest and most massive in a family of bolometric detectors designed to search for the \bbz decay of \tect(Q-value= \unit[2528]{keV}~\cite{Frank,Nick,Rahaman2011412}). The detector combines the excellent energy resolution achievable with the bolometric technique (\ca~5~keV at 2615~keV) with the exceptionally high natural abundance of \tect (34\%). The first phase of CUORE, named CUORE-0, was \allowbreak 1/19$^{th}$ the size of CUORE and operated at the Laboratori Nazionali del Gran Sasso (LNGS), in Italy, between 2013 and 2015. In addition to being a competitive \bbz\allowbreak experiment~\cite{Q0Preliminary,Us,Q0-analysis}, \qz is a test of the assembly protocols for CUORE: the reconstruction of background sources responsible for the \qz counting rate enables us to verify that the necessary background requirements for CUORE are fulfilled. 

In bolometers, the neutrinos emitted in \bbd are not detected, and the summed kinetic energy of the two electrons forms a continuous spectrum from \unit[0]{keV} up to the Q-value of the decay. Conversely, \bbz produces no neutrinos and the experimental signature is a sharp peak at the Q-value of the decay, broadened by the energy resolution of the detector. This broadening smears \bbd events into the \bbz region of interest (ROI) around the Q-value and forms an irreducible background to the \bbz signal; thus a good energy resolution is key to mitigating this background.

The other background contributions in the ROI come from naturally occurring radioactivity in the detector components. These background sources can be disentangled and described quantitatively by carefully analyzing the shape of the measured spectrum and constructing a detailed background model, including both physics processes and instrumental effects.

This paper reports the CUORE-0 background model as well as a new precision measurement of the \tect \bbd half-life. We use a detailed Geant4-based simulation and a Bayesian fitting algorithm with \textit{a priori} constraints from materials assay to reconstruct the experimental data and form \textit{a posteriori} estimates of the background source activities. A frequentist analysis, using the same model, is presented in~\cite{Brian}. 

We present the experimental details, including the data acquisition and analysis chain, in Sec.~\ref{sec:exp}. Background constraints from external measurements are summarized in Sec.~\ref{sec:sources}. 
The construction of the Monte Carlo simulation code is presented in Sec.~\ref{sec:MC}. 
The validation of Monte Carlo simulations, by comparing external radioactive source calibration spectra with the data, is presented in Sec.~\ref{sec:calibration}. 
The set of identified background sources and their effects on the experimental data is discussed in Sec.~\ref{sec:bkg_model}. The Bayesian fitting tool is introduced in Sec.~\ref{sec:jags}. We present the fit results and discuss their uncertainties in Sec.~\ref{sec:systematics}. Finally, the \bbd decay half-life evaluation is presented in Sec.~\ref{sec:results} and the \bbz ROI reconstruction is discussed in Sec.~\ref{sec:ROI}.

%% file: exp.tex
The CUORE-0 detector is one tower of 52 $^{\mathrm{nat}}$TeO$_2$ crystals arranged in 13 floors, each with 4 crystals \cite{Q0-detector}. Each crystal is \unit[750]{g} and is operated as an independent bolometer at $\sim$\unit[10]{mK}. At these temperatures, the interaction of a particle with the crystal generates a measurable temperature rise proportional to the energy deposited. The total detector mass is \unit[39]{kg} of \teod, or \unit[10.8]{kg} of \tect.

\begin{figure*}
\begin{center}
\includegraphics[width=15cm]{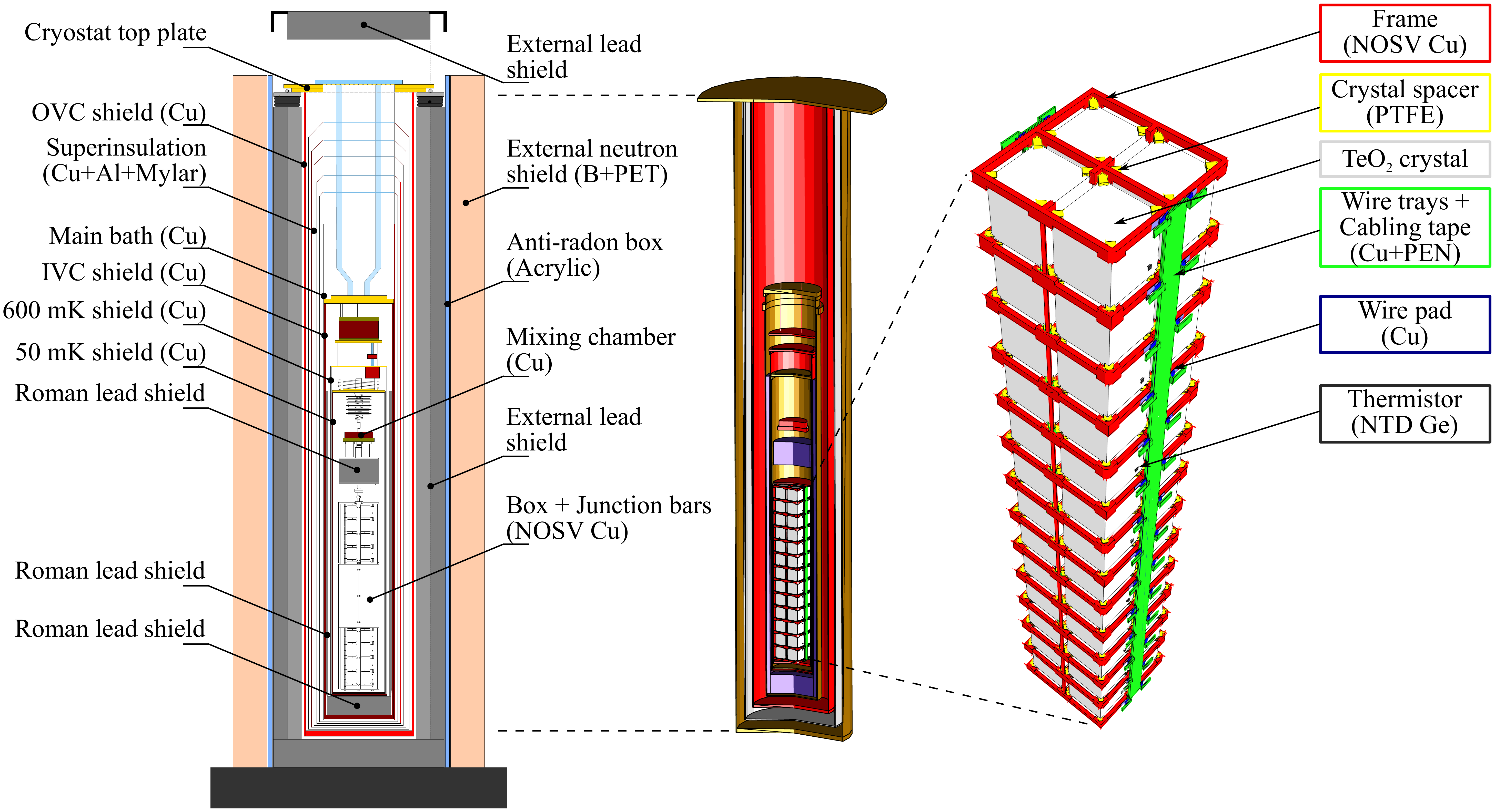} %
\caption{The \qz as built (Left) compared to the implementation in the simulation (Middle). The details of the tower structure are shown as implemented in the simulation (Right). Detector floors are numbered starting with floor 1 at the bottom.}
\label{fig:cuore0setup}
\end{center}
\end{figure*}

The tower is situated in a dilution refrigerator that provides the cooling power necessary to keep the \teod crystals at their working temperature. The copper and PTFE support structure holds the crystals and provides the thermal link to the refrigerator. The tower is surrounded by several layers of shielding, including low-background Roman lead and an anti-radon box. A schematic of the experiment is shown in Fig.~\ref{fig:cuore0setup} (Left).

The crystals temperatures are measured continuously using neutron transmutation doped (NTD) germanium thermistors, coupled to the crystals using epoxy. The thermistors convert temperature variations to a voltage output, which is digitized at a rate of \unit[125]{Samples/s}. A software trigger is used to identify events and collect them in \unit[5]{s} windows. Each window is divided into two periods: \unit[1]{s} before the trigger and \unit[4]{s} after it. The period before the trigger is used to establish the baseline temperature of the crystal and the remaining \unit[4]{s} is used to determine the pulse amplitude. Together these are used to extract the energy deposited. 

A silicon resistor is also coupled to each crystal with epoxy and is used to generate reference thermal pulses every \unit[300]{s}. These are used to stabilize the gain of the bolometer against temperature fluctuations. Forced triggers are used for noise and threshold studies. 
The threshold (in the default high-energy triggered data used for this analysis) is bolometer dependent and is approximately between \unit[30] and \unit[120]{keV}.
The energy response is calibrated by inserting thoriated tungsten wires inside the outer vessel of the cryostat, and using the \gm lines from the \ths decay chain to calibrate each bolometer independently.

The details of the CUORE-0 detector design, operation and performance are described in~\cite{Q0-detector}. New protocols for material selection, cleaning, and handling were developed for the detector crystals and tower support structure. The dilution refrigerator, shielding, and other cryostat components are those from the \qino experiment~\cite{qino-2008,qino-2011}.

\subsection{Data Production}
This analysis uses data collected with CUORE-0 from March 2013 to March 2015. The data are grouped in datasets, which last approximately one month. Each dataset has approximately three days of calibration data at both the beginning and end, collected while the detector was exposed to a radioactive calibration source.  
In total, 20 datasets were used in the \bbz analysis: however, in the present analysis we exclude the first dataset because of a substantial ($\sim$15\,keV) miscalibration in the highest energy region (above 3 MeV). The physics data (i.e. excluding the calibration data of each dataset) use a total of \unit[33.4]{kg$\cdot$y} of \teod exposure, or \unit[9.3]{kg$\cdot$y}  of \tect exposure, after all data quality selections. 

The data production converts the data from a series of triggered waveforms into a calibrated energy spectrum. The details of the data production can be found in~\cite{Q0-analysis}, but we outline the general procedure here. The entire waveform is used to extract the amplitude of the pulse. Then, each bolometer in each dataset is calibrated  independently. A time-coincidence analysis is performed to search for events that deposit energy across multiple bolometers. Finally, the \bbz ROI (2470-2580 keV) is blinded for analysis. 

Once the data are blinded, we implement a series of event selection cuts to maximize our sensitivity to physics events. We exclude periods of cryostat instability and malfunction. We enforce a pile-up cut of \unit[7.1]{s} around each event: the \unit[3.1]{s} before and \unit[4]{s} after the event. We apply a series of pulse shape cuts to reject deformed or non-physical events mostly contaminating the low energy region below $\sim$\unit[400]{keV}.

Double-beta decay events are usually confined within the crystal they originated from. However, many background sources deposit energy in multiple crystals within the response time of the detector. We use this information in the analysis by forming multiplets of events that occur within a coincidence window of $\pm$\unit[5]{ms} in different crystals. Since the event rate is approximately 1~mHz, the probability of accidental (i.e. causally unrelated) coincidences is extremely small ($\sim 10^{-5}$). 

We then build energy spectra from these multiplets:
\begin{itemize}
\item{\bf \mspec} spectrum is the energy spectrum of all events, with no coincidence criteria applied;
\item{\bf \mspecone} spectrum is the energy spectrum of the events with the requirement that only one bolometer triggered (multiplicity 1 or {\bf \mspecone} events);
\item{\bf \mspectwo} spectrum is the energy spectrum of the events with the requirement that two bolometers triggered (multiplicity 2 or {\bf \mspectwo} events);
\item{\bf \summspectwo} spectrum is the energy spectrum associated to \mspectwo multiplets, each multiplet produces an entry with an energy $E($\summspectwo$)$ that is the sum of the energies of the two events composing the multiplet.

\end{itemize}

Higher-order multiplets are used only to evaluate the contribution of muons to the background.

The signal cut efficiency as a function of energy is defined as the fraction of true signal events that pass all the event cuts.
In~\cite{Q0-analysis}, we calculated the efficiency of these cuts by measuring their effect on \gm-peaks in the energy spectrum. This takes advantage of the fact that the events in the \gm-peaks are a nearly pure sample of true signal events. In the present analysis we use a new technique, which takes advantage of the coincidence analysis and allows a better reproduction of the energy dependence of the efficiency. Since accidental coincidences are negligible, \mspectwo events provide a pure sample of good events on the whole energy spectrum. The cut efficiency $\varepsilon_C$ is modeled as an exponential function of the energy, $\varepsilon_C(1-e^{-E/E_c})$, which is fitted to \mspectwo events on a bin-by-bin basis; it rapidly reaches a stable value of $\varepsilon_C = 0.943\pm0.002$ at energies above $ E_c\sim 100\,\mathrm{keV}$. This is consistent with the signal cut efficiency quoted in~\cite{Q0-analysis} of $0.937\pm0.007$. Above $E\sim 7\,\mathrm{MeV}$ the {\mspectwo} spectra have insufficient statistics to give meaningful fit results. Therefore, in this analysis we do not include events with energy above \unit[7]{MeV}.

%% file: bkg_sources.tex
In rare event searches, some backgrounds are ubiquitous due to the natural decay of \kq and the daughters of the \ths and \u decay chains, including the surface implantation of \pbdd from environmental \rndd. Based on the results of previous bolometric experiments~\cite{qino-2008,qino-2011,EPJ2}, we expect 
these radioactive contaminants to be located in the whole experimental setup, including the detector itself. The cosmogenic activation, especially of copper and tellurium, resulting in \cosz is also of concern.
A small contribution from cosmic muons~\cite{CuoricinoMuons}, environmental gamma rays (\gm)~\cite{CuoreExternal}, and neutrons interacting directly in the detector is expected~\cite{CuoreExternal}.
With this in mind, extreme care  was taken into the selection of the materials used to build the \qz\ detector \cite{Q0-detector}, and in the cleaning of all surfaces facing each \teod\ crystal \cite{TTT,Arnaboldi:2010cw}.  
A lot of effort was also devoted to: the detector design, to minimize both the total mass of the inactive parts of the detector tower and the surface area facing the array; the detector assembly procedure under controlled atmosphere \cite{Q0-detector}; and the optimization of the production protocol of every detector component to limit exposure to cosmic rays.

Tables~\ref{tab:bulk} and~\ref{tab:surf} report the bulk and surface activities from the screening of the \qz components. A discussion about the different assay techniques that we used to derive the quoted contamination limits can be found in~\cite{EPJ2,TTT}. The \qino results (this experiment used the same cryostat and shield as CUORE-0) and screening results guide the construction of the background model and the definition of the \emph{priors} on material contaminants.

%
\begin{table}[!htb]
\begin{center}
\caption{
Measurements and limits on bulk contaminations of the various detector components, as obtained with different measurement techniques: bolometric, Neutron Activation Analysis, Inductively Coupled Plasma Mass Spectrometry, 
High Purity Ge 
\gm spectroscopy. Error bars are 1 sigma, limits are 90\% C.L. upper limits.}
\begin{tabular}{cccc}
\hline
Component                  & $^{232}$Th                  & $^{238}$U                  & $^{40}$K         \\
                           & [Bq/kg]                     & [Bq/kg]                    & [Bq/kg]          \\
\noalign{\smallskip}\hline
TeO$_2$ crystals            & $<$8.4$\cdot$10$^{-7}$     &  $<$6.7$\cdot$10$^{-7}$    & \\
Epoxy                        & $<$8.9$\cdot$10$^{-4}$     &  $<$1.0$\cdot$10$^{-2}$    &$<$47$\cdot$10$^{-3}$ \\
Au bonding wires            & $<$4.1$\cdot$10$^{-2}$     &  $<$1.2$\cdot$10$^{-2}$    &   				\\
Si heaters                  & $<$3.3$\cdot$10$^{-4}$     &  $<$2.1$\cdot$10$^{-3}$    &   						\\
Ge thermistors          & $<$4.1$\cdot$10$^{-3}$     &  $<$1.2$\cdot$10$^{-2}$    &     					\\
PEN-Cu cables               & $<$1.0$\cdot$10$^{-3}$     &  $<$1.3$\cdot$10$^{-3}$    &$<$1.3$\cdot$10$^{-2}$   \\ 
PTFE supports               & $<$6.1$\cdot$10$^{-6}$     &  $<$2.2$\cdot$10$^{-5}$    &   						\\
Cu NOSV                     & $<$2.0$\cdot$10$^{-6}$     &  $<$6.5$\cdot$10$^{-5}$    &7$\pm$2$\cdot$10$^{-4}$  \\
Pb Roman                    & $<$4.5$\cdot$10$^{-5}$	 &  $<$4.6$\cdot$10$^{-5}$    &$<$2.3$\cdot$10$^{-5}$   \\ 
Pb Ext                      & $<$2.6$\cdot$10$^{-4}$	 &  $<$7.0$\cdot$10$^{-4}$    &$<$5.4$\cdot$10$^{-3}$     \\ 

\noalign{\smallskip}\hline
\end{tabular}
\label{tab:bulk}
\end{center}
\end{table}
\begin{table}[!htb]
\begin{center}
\caption{90\% C.L. upper limits for the surface contaminants of the most relevant elements facing the \qz\ detector, as obtained with different measurement techniques: bolometric, Neutron Activation Analysis, and \alph spectroscopy with Si barrier detectors. Different contamination depths are considered: 0.01-10\,$\mu$m for crystals; 0.1-10\,$\mu$m for heaters, thermistors and CuNOSV; and 0.1-30\,$\mu$m for PEN and PTFE components (contamination depths are further discussed in Sec.\,\ref{sec:alpha}).}
\begin{tabular}{cccc}
\hline
Component     & $^{232}$Th & $^{238}$U & $^{210}$Pb\\
           &[Bq/cm$^{2}$]& [Bq/cm$^{2}$]& [Bq/cm$^{2}$]\\
\noalign{\smallskip}\hline
TeO$_2$ crystals~\cite{ccvr}       	& $<$2$\cdot$10$^{-9}$          & $<$9$\cdot$10$^{-9}$          & $<$1$\cdot$10$^{-6}$\\
Si heaters     \cite{rad2005,rad2006}    		& $<$3$\cdot$10$^{-6}$          & $<$8$\cdot$10$^{-7}$          & $<$8$\cdot$10$^{-7}$\\
Ge thermistors 						& $<$8$\cdot$10$^{-6}$        	& $<$5$\cdot$10$^{-6}$        	& $<$4$\cdot$10$^{-5}$\\
PEN-Cu cables       				&$<$4$\cdot$10$^{-6}$          & $<$5$\cdot$10$^{-6}$          & $<$3$\cdot$10$^{-5}$\\
PTFE supports           	& $<$2$\cdot$10$^{-8}$        & $<$7$\cdot$10$^{-8}$        & \\
CuNOSV    \cite{TTT}               	& $<$7$\cdot$10$^{-8}$        	& $<$7$\cdot$10$^{-8}$        	& $<$9$\cdot$10$^{-7}$\\  
\noalign{\smallskip}\hline     
\end{tabular}
\label{tab:surf}
\end{center}
\end{table}

%% file: montecarlo.tex
The background sources are simulated using a Geant4-based Monte Carlo code called \mcuorez.  The code generates and propagates primary and any secondary particles through the CUORE-0 geometry until they are detected in the \teod \allowbreak crystals. The code outputs the energy and time of the energy depositions (time is used to properly take into account correlations in nuclear decay chains). A second program takes the output of \mcuorez and applies a detector response function and incorporates other read-out features.
 
\subsection{Monte Carlo Simulation}
\mcuorez is implemented in Geant4 version 4.9.6.p03. $\alpha$, $\beta$ and $\gamma$ particles, nuclear recoils, neutrons and muons are propagated down to keV energies, with an optimization done on the different volumes to balance simulation accuracy and speed.
We have chosen the Livermore physics list, and particles can be generated and propagated in the bulk and on the surface of all components. The surface contamination is modeled, according to diffusion processes, with an exponential density profile and a variable depth parameter.

\sloppy Single radioactive decays as well as the \u and \ths decay chains have been implemented using the G4RadioactiveDecay database. \bbd is parameterized according to~\cite{Kotila}. The generation of external muons, neutrons and \gm is described in~\cite{CuoricinoMuons,CuoreExternal}.

\begin{table*}[t]
\begin{center}
\caption{Elements of the \qz setup implemented in \mcuorez with the values of their volumes, surface and mass. Inside the cryostat all the volumes are in vacuum but the Main Bath that is filled with LHe. In Column (5) we indicate the short-name of the component that is used in this paper. Some components are not used as source contamination in this analysis, but they exist in the simulation, including their absorption properties. }
\begin{tabular}{lcccc}
\hline
{\bf Component Description}        	      &{\bf Volume [dm$^3$]}	&{\bf Surface [dm$^2$]}    &{\bf Mass [kg]}    & {\bf Component Name}     \\

 \hline
 External Neutron Shield (B+PET)	&	1385.2	&	2859.2	&	2770.3	& \\
 \hline
 External Lead Shield	(Pb) &	2177.6	&	2338.9	&	24694.0	& \pbext \\
\hline 
 OVC Shield	(Cu) &	53.2	&	894.6	&	474.8	& \\
 Superinsulation layers	(Cu+Mylar+Al)&	18.3	&	728.0	&	163.2	& \cryoext \\ 
 Main Bath	(Cu) &	16.3	&	692.7	&	145.0	& \\
\hline

 Cryostat Top Plate (Brass) &	0.3	&	6.9	&	2.3 &	\\
 \hline
 Dilution Unit (Fe) &	0.5	&	4.3	&	3.8	&  \\ 
\hline
 IVC Shield	(Cu) &	5.1	&	278.3	&	45.6	&  \\
 600 mK	Shield (Cu) &	2.2	&	178.7	&	19.8	& \cryoint \\
 50 mK	Shield (Cu) &	1.9	&	154.8	&	16.9	&  \\
\hline
 Roman Lead Shield	&	17.8	&	198.7	&	202.3	& \pbint \\
\hline 
 Detector Tower Box	(NOSV Cu) &	0.8	&	93.3	&	6.8	& \\
 Frames (NOSV Cu)	&	0.3	&	23.5	&	2.6	& \holder \\
 Wire trays	(NOSV Cu)&	0.1	&	14.4	&	0.6	& \\
 Junction bars (NOSV Cu)&	0.1	&	14.4	&	0.6	& \\
\hline 
 NTD thermistors (Ge) &	4.7 $\times$ 10$^{-4}$	&	0.2	&	0.002	& \\
 Cabling tapes (Cu+PEN)	&	2.2 $\times$ 10$^{-2}$	&	5.8	&	0.04	& \smallpart \\
 Wire pads	(Cu)&	3.1$\times$ 10$^{-4}$	&	0.6	&	0.003 &   \\
 Crystals spacers (PTFE)	&	0.1	&	12.6	&	0.2	& \\
\hline
 \teod crystals	&	6.5	&	78.0	&	39.0	& \crystal \\ 
\hline	
\end{tabular}
\label{tab:geometry}
\end{center}
\end{table*}

Due to the low counting statistics in CUORE-0 data some components are indistiguishable (i.e. exhibit degenerate spectral shapes)\footnote{Some degenerate sources could be disentangled by considering distribution of events across the detector. In order to keep the number of free parameters small and the number of simulated configurations tractable, we do not use position of the events in this analysis, and average the spectra over the entire tower.} and can be grouped, provided that the prior on their material contamination is properly evaluated. This simplification reduces the number of free parameters in the final fit of the simulation to the data. A similar case holds for components made of the same material (i.e., characterized by an identical contamination): their simulations can be grouped once scaled by mass or surface to properly equalize the contamination densities.

The detector as built and the geometry as implemented in \mcuorez are shown in Fig.~\ref{fig:cuore0setup}. The outermost volume included in CUORE-0 geometry is the external neutron shield. Although its contamination is negligible, its effect on the propagation of external neutrons and muons is important.

The next layer is the external lead shield made of modern lead (\pbext). 
Only the inner \unit[10]{cm} are simulated, since self-shielding is so high that the contribution from the outer volume is negligible.    

The cryostat is then modeled as composed by two volumes: the Cryostat External Shields (\cryoext) and the Cryostat Internal Shields (\cryoint). 

The \cryoext groups three components made of copper that has been underground for  more than 25 years. These components exhibit degenerate spectra and the only prior considered for their contamination is that on $^{60}$Co derived from their identical history. Included in this volume are a small amount of superinsulation material (Mylar and Al) and the gap between the OVC (cryostat Outer Vacuum Chamber) and the \pbext. The \pbext  is enclosed in acrylic glass and flushed with nitrogen (anti-radon box) to prevent radon from filling this gap and entering the experimental volume. 

The \cryoint groups three shields made of the same newer copper, underground since 2002: the IVC (cryostat Inner Vacuum Chamber), the 600 mK and 50 mK shields. The roman lead ~\cite{RmLead} shield (\pbint) is inserted between the IVC and the 600 mK shield. The \holder is the structure that holds the crystals. It is made from NOSV copper, a special copper alloy from Aurubis company suitable for cryogenic use, cleaned according to the CUORE protocols~\cite{Q0-detector}. It has two main parts: the frame that supports the crystals and the surrounding cylindrical box. Finally, the \crystal are designed as \teod cubes with identical contaminations.

In Table~\ref{tab:geometry}, the components listed as \smallpart are not included in the following analysis because of their small mass and negligible contamination (see Tables~\ref{tab:bulk} and \ref{tab:surf}). Only the PTFE spacers could provide a sizable contribution to the background, however their spectra are degenerate with the \holder ones, therefore their contribution is included in the latter element.

\subsection{Monte Carlo Data Production}
In order to make the Monte Carlo reproduce experimental data - as already anticipated - a second code is used to recreate the detector time and energy response. 
We account for the timing resolution of each crystal by combining energy depositions that occur in the same crystal within a window of $\pm$\unit[5]{ms}. The absolute time of events is assigned based on a random distribution with an event rate of \unit[1]{mHz} (the experimental rate during physics runs) and any physical coincidences from \mcuorez are preserved. 

Once the simulated events are correlated correctly in time, the resulting energy depositions are smeared with a Gaussian energy response function. The width of the function varies linearly with energy and is based on the FWHM resolution measured on \gm peaks in the \mspec spectrum, between 511 and \unit[2615]{keV}. 

We reproduce also the energy dependence of the trigger efficiency. This is experimentally evaluated for each bolometer using heater-generated pulses with variable amplitudes. For each amplitude (then converted into a particle equivalent energy) the efficiency is defined as the number of triggered signals over the generated ones.  The efficiency vs. energy curve is interpolated with an Erf function and used in simulation data production.

As is done in the experiment, events with coincidences in multiple bolometers within a $\pm$\unit[5]{ms} window are combined into multiplets. Pile-up events (i.e. events occurring in the same bolometer within the pile-up rejection window set by the analysis -- see Section~\ref{sec:exp}) are removed to account for dead time. 

The simulation properly reproduce the \gm particle energies, since we calibrate CUORE-0 spectra using \ths \gm lines. However, we observe in our data an energy quenching effect for \alph particles compared to \gm, which makes the measured energies higher than the known energies~\cite{alpha1,alpha2}. We account for this effect in the simulations by shifting \alph energy depositions by 0.8\%, which is the average shift observed for \ths and \u \alph peaks visible in CUORE-0 background spectrum.

%% file: calibration.tex
\begin{figure*}[thbp!]
\includegraphics[width=.99\textwidth]{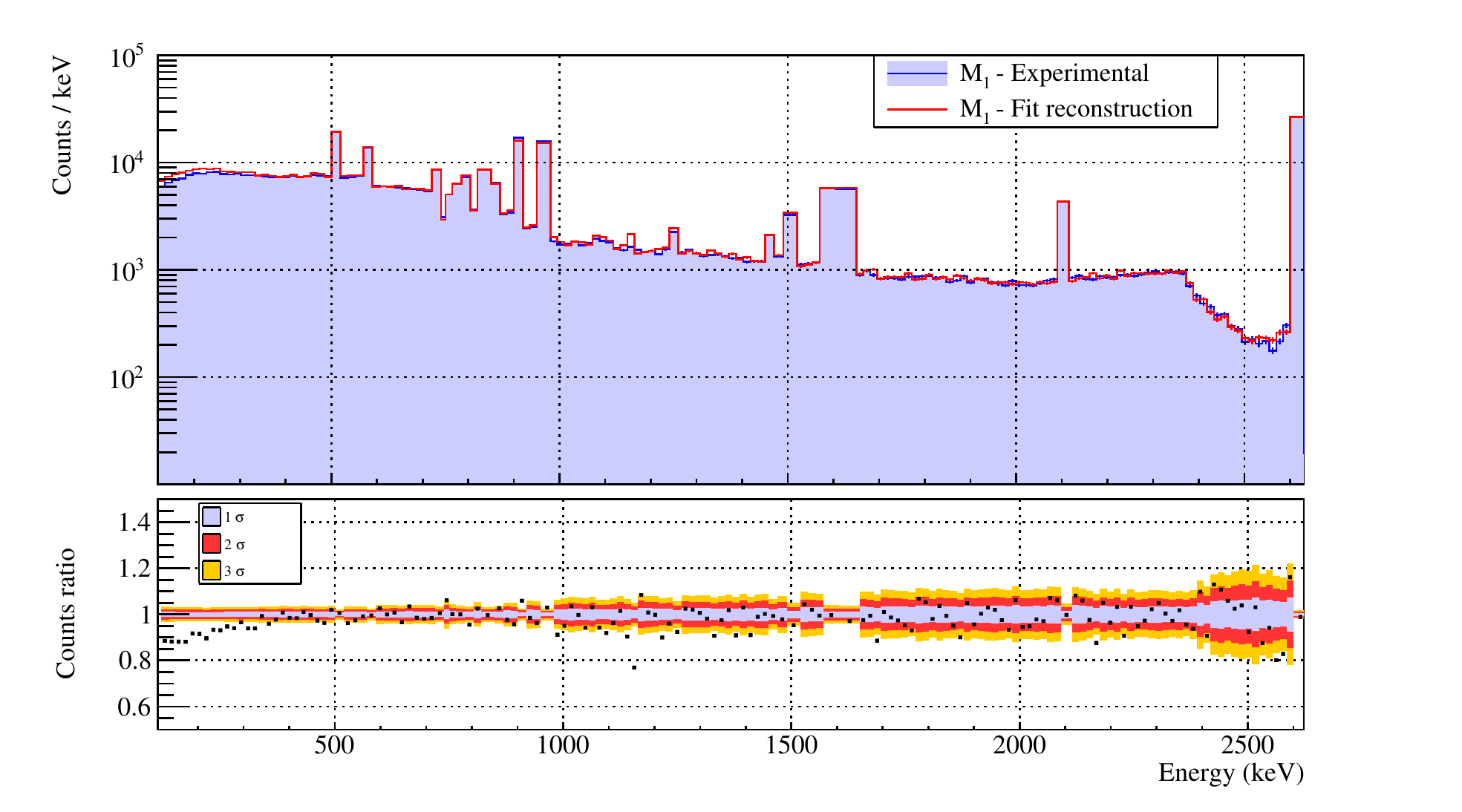}   
\caption{CUORE-0 source calibration measurement. Top panel: Comparison of \mspecone spectrum with its \mcuorez simulation. Peaks are grouped in a single bin, while in the continuum the minimum bin size is \unit[15]{keV}. Bottom panel: Bin-by-bin ratio between counts in the experimental spectrum and counts in the reconstructed one. We observe a less than 10\% discrepancy in the rate at low energies 
	($<300$\,keV) that is more likely due to small errors in the geometry reconstruction.} 
\label{fig:calibrazione}
\end{figure*}

We use calibration data to test the \mcuorez simulation. The thoriated wire calibration source is deployed between the \cryoext and the \pbext, therefore its simulation involves all volumes internal to the \pbext. The calibration data rate is about 100 times higher than in physics runs, so pile-up effects become sizeable. Fig.~\ref{fig:calibrazione} shows that with proper treatment of pile-up effects, we achieve a good data-simulation agreement. We observe small deviations in the low energy region between 100 and 300 keV and on a few peaks. The deviation in the low energy region (the origin is still unknown) is a potential source of systematic errors that will be considered in Sec.~\ref{sec:systematics}. For what concerns peaks, the largest deviation observed is a 7\% difference on the \acddo line at 1153 keV (branching ratio 0.1\%). We traced this effect to the Geant4 version\footnote{We recently verified that this problem is gone in Geant4 version 4.10.}; anyway it has no consequences on the background reconstruction since it involves only few very low intensity lines.
 
%% file: bkg_model.tex
\begin{figure*}[htbp!]
	\begin{center}
   \includegraphics[width=1\textwidth]{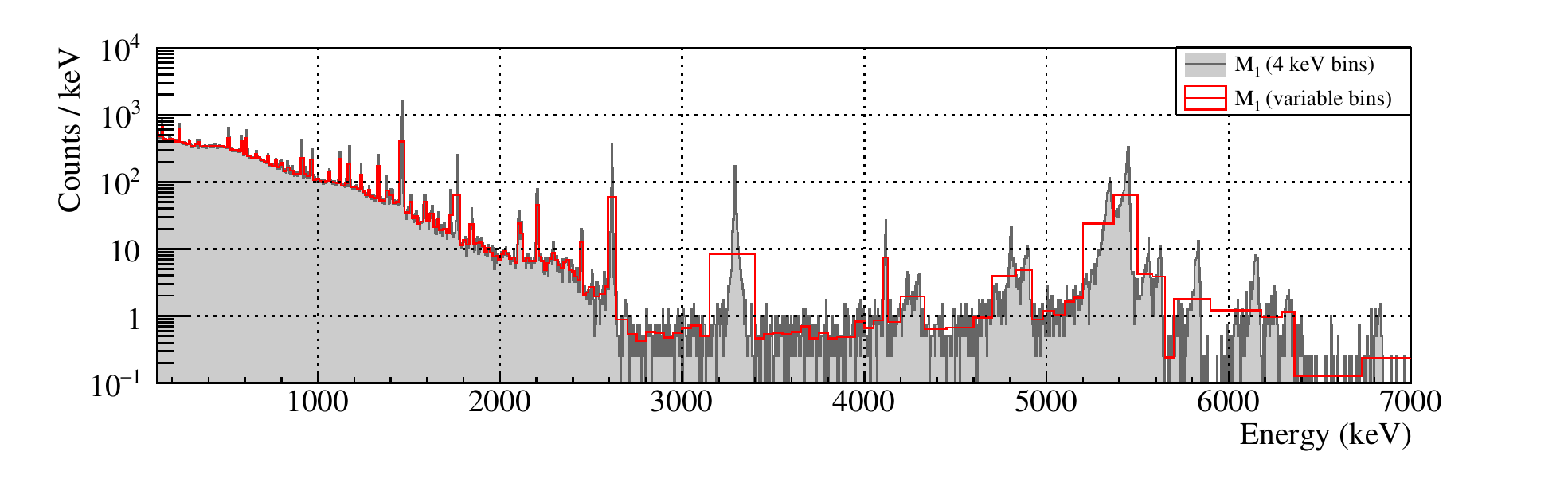}
	\end{center}
	\caption{CUORE-0 \mspecone spectrum with fixed 4 keV binning (gray) and with the variable binning used for the background model fit (red). The variable binning is used to reduce the effects of statistical fluctuation and/or of line shape (see text for more details).}
	\label{fig:M1-Spectrum} 
\end{figure*}

The really critical part of the  
background model construction is the  determination \emph{a priori} of the most relevant sources to be included in the fit. Omitting a relevant source could lead to a poor fit or, even worse, a good fit to the wrong model. The degeneracy of many of the sources energy spectra introduces a further complication.
The analysis reported in this section uses lines in the \mspecone and \summspectwo spectra to identify the radioactive isotopes contributing to the background counting rate, their possible location and any other characteristic feature that needs to be considered in the construction of the background model. 

Sources such as the external muons, neutrons, and gammas  are straight-forward to model since their fluxes are well known from independent laboratory measurements. The radioactive decays are more difficult to disentangle. The location and distribution of the contaminants must be specified, and in the case of decay chains a choice must be made between the secular equilibrium hypothesis and a break in the chain due to material handling. We use the external screening measurements presented in Sec.\,\ref{sec:sources}, estimates of cosmogenic exposure, and distinctive features in the data itself to select the most probable sources. These data also provide the priors for the fit. 
 
Figure~\ref{fig:M1-Spectrum} shows the \mspecone spectrum. The spectrum below the \unit[2.615]{MeV} \gm line of $^{208}$Tl includes many \gm lines. Above \unit[2.615]{MeV} it is dominated by \alph\allowbreak events. These are the \gm and \alph regions respectively. The energies of the \gm and \alph lines, the time variation of their counting rates, and the observation of prompt or delayed coincidences can all be used to select the final list of sources to be included in the fit.

\subsection{CUORE-0 \gm Region Analysis}\label{sec:gamma}

\input{gamma.tex}

\subsection{CUORE-0 \alph Region Analysis}\label{sec:alpha}
\input{alfa.tex}

\subsection{Source list}\label{sec:list}
\begin{figure}[tbp]
	\begin{center}
	\includegraphics[width=0.45\textwidth]{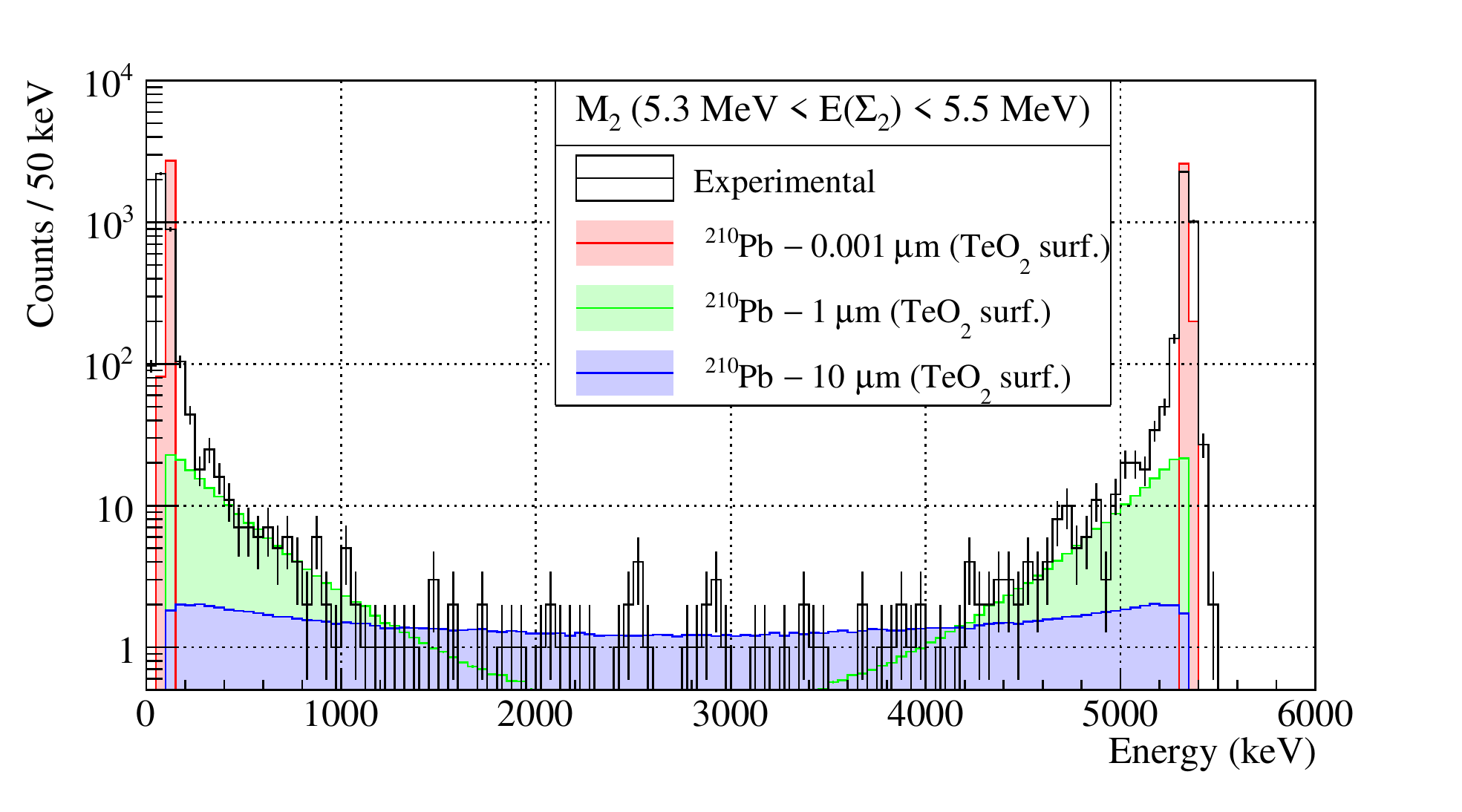}
	\end{center}
	\caption{Comparison between the experimental and the Monte Carlo \mspectwo with \summspectwo energy equal to the Q-value of \podd \alph decay. The MC spectra refer to \crystal surface contamination with different depths parameters: 0.001, 1, and \unit[10]{\ums}. The normalization here is chosen to approximately reproduce the profile of the experimental data.}	
	\label{fig:alphaprofile} 
\end{figure}

A list of the 57 sources used for the present analysis is shown in Table~\ref{tab:SourceList}. The components in the first column are the 6 defined in the fifth column of Table~\ref{tab:geometry}. The priors in the fourth column derive mainly from Table~\ref{tab:bulk}. The priors for the \cosz activities are discussed in Sec.~\ref{sec:gamma}. 
There are two components included in the analysis but not included in Table~\ref{tab:geometry}. One is the \emph{\unit[50]{mK} Spot}, a point-like source located on the internal surface of the \unit[50]{mK} thermal shield facing tower floor~10 (see Fig~\ref{fig:cuore0setup}). The other is the \emph{Bottom Plate} \pbext, a disc-like source placed on the internal bottom plate of the external lead shield \pbext. 
These two sources model the \ths excess observed on the floor~number 10 and the \kq excess observed on bottom floors, which are discussed in Sec~\ref{sec:gamma}.
In order to properly reproduce the shape of the \alph peaks in \mspecone and \mspectwo spectra, \crystal and \holder surface contamination with a few representative depths are included in the analysis (Fig.~\ref{fig:alphaprofile}). The exact choice of contaminant depth is treated as a systematic uncertainty. This is discussed in Sec.~\ref{sec:systematics}. 

Violations of secular equilibrium in the \u and \ths decay chains are only considered when they produce a distinguishable signature in the spectrum. 
A background contribution due to muons is included in the analysis; contributions from neutrons and external $\gamma$s are negligible. Muons contribute mainly through $\gamma$s produced by interactions in the detector components. 
A prior is obtained from high multiplicity data, which is compatible with the muon flux measured by other experiments.

%% file: gamma.tex
\begin{figure}[htb]
	\begin{center}
		\includegraphics[width=.49\textwidth]{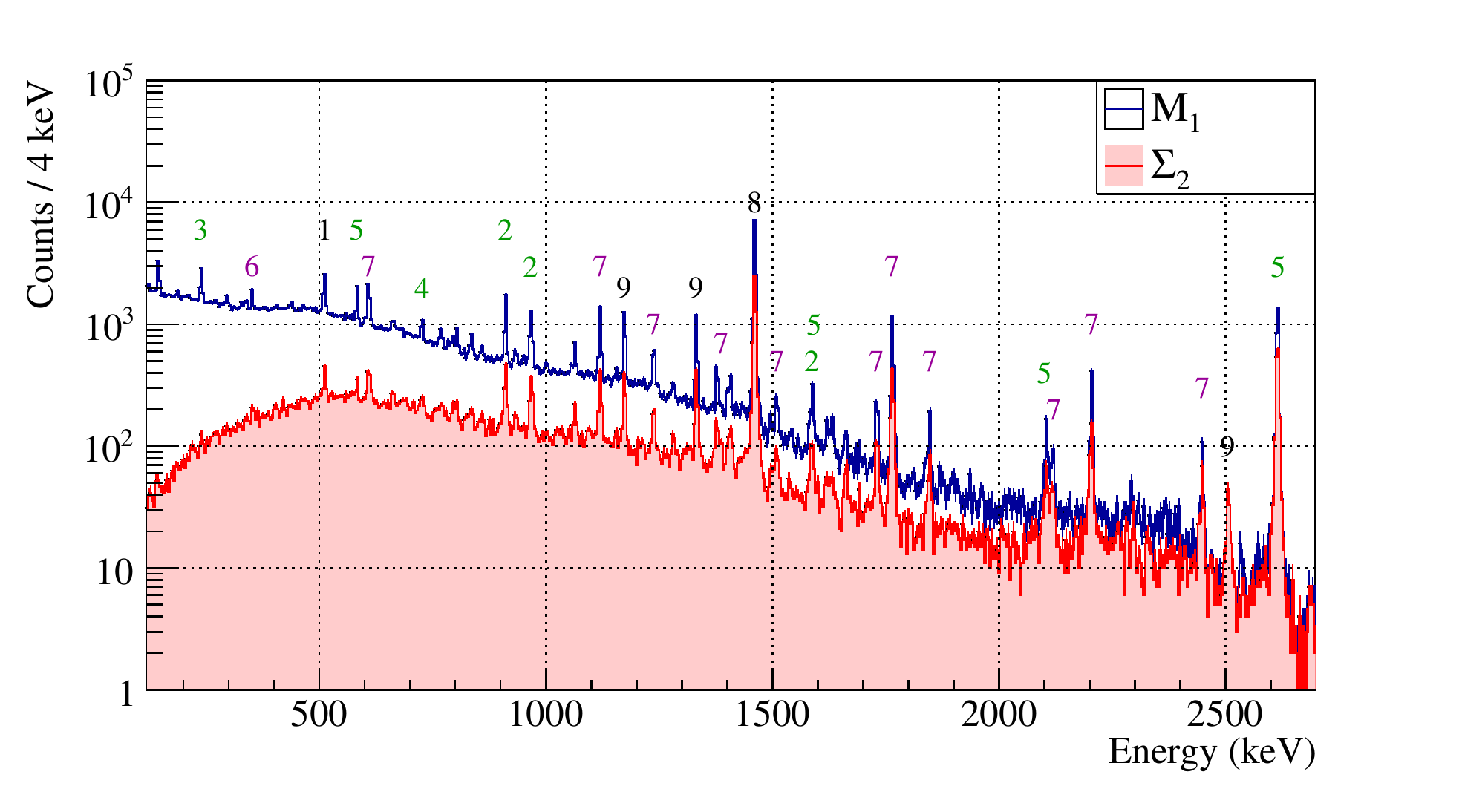}
		\caption{CUORE-0 \mspecone (blue) and \summspectwo (red) in the \gm region. The peaks are labeled as follows: (1) $e^+e^-$ annihilation, (2) $^{228}$Ac, (3) $^{212}$Pb, (4) $^{212}$Bi, (5) $^{208}$Tl, (6) $^{214}$Pb, (7) $^{214}$Bi, (8) $^{40}$K, (9) $^{60}$Co.}
		\label{fig:g_region}
	\end{center}
\end{figure}

\ths and \u are natural long-lived radionuclides that generate radioactive decay chains. They can be found in almost all materials. The lines from the \ths and \u decay chains are clearly visible in Fig.~\ref{fig:g_region}; their intensities measured by fitting the \mspec spectrum are reported in Tables~\ref{tab:th-lines} and \ref{tab:u-lines}. The lines from \kq and other contaminants are also present; see Table~\ref{tab:gamma-lines}.

\begin{table}[!htb]
\begin{center}
\caption{\gm lines belonging to \ths chain observed in \qz spectrum. Only lines with branching ratios above 1\% are reported.}
\begin{tabular}{rcr}

\hline
Energy 		& Isotope	& Rate		\\
(keV)	& & {(counts/kg/y)}   \\
\hline
238.6		& \pbdud		& 51.0   $\pm$ 3.5   \\
300.1		& \pbdud		& 2.3  $\pm$ 1.7		\\
328.0		& \acddo		& 6.4  $\pm$ 1.7		\\
338.3		& \acddo		& 13.5  $\pm$ 2.1	\\
583.2		& \tld			& 44.0  $\pm$ 2.3	\\
727.3		& \bidud		& 12.6  $\pm$ 1.5	\\
795.0		& \acddo		& 8.4  $\pm$ 1.4		\\
860.6		& \tld			& 8.0  $\pm$ 2.3		\\
911.3		& \acddo		& 49.2  $\pm$ 1.8	\\
964.8		& \acddo		& 11.3  $\pm$ 1.2	\\
969.0		& \acddo		& 29.7  $\pm$ 1.6	\\
1588.1		& \acddo		& 14.9  $\pm$ 1.2	\\
1620.5		& \bidud		& 2.9  $\pm$ 0.7		\\
1630.6		& \acddo		& 3.9  $\pm$ 0.8		\\
2614.5		& \tld			& 80.9  $\pm$ 2.3	\\

\noalign{\smallskip}\hline
\end{tabular}
\label{tab:th-lines}
\end{center}
\end{table}
\begin{table}[!htb]
\begin{center}
\caption{\gm lines belonging to \u chain observed in \qz\ spectrum. Only lines with branching ratios above 1\% are reported. The peaks at 242\,keV and 786\,keV include a contribution from \raddq and \bidud of the \ths chain, respectively.}
\begin{tabular}{rcr}

\hline
Energy 		& Isotope	& Rate		\\
(keV)	& & (counts/kg/y)   \\
\hline
242.0	& \pbduq	& 6.5 $\pm$ 2.5  	 \\
295.2		& \pbduq		& 10.6 $\pm$ 2.3  	 \\
351.9		& \pbduq		& 20.9 $\pm$ 2.7  	 \\
609.3		& \bidq		& 56.8 $\pm$ 2.3  	 \\
768.4		& \bidq		& 7.1 $\pm$ 2.1  	 \\
786.0	& \pbduq	& 3.9 $\pm$ 1.4  	 \\
934.1		& \bidq		& 5.6 $\pm$ 1.1  	 \\
1120.3		& \bidq		& 41.4 $\pm$ 1.6  	 \\
1155.2		& \bidq		& 4.4 $\pm$ 1.0  	 \\
1238.1		& \bidq		& 17.4 $\pm$ 1.3  	 \\
1281.0		& \bidq		& 4.9 $\pm$ 1.5  	 \\
1377.7		& \bidq		& 12.3 $\pm$ 1.1  	 \\
1401.5		& \bidq		& 5.7 $\pm$ 0.9  	 \\
1408.0		& \bidq		& 7.9 $\pm$ 1.0  	 \\
1509.2		& \bidq		& 7.5 $\pm$ 1.0  	 \\
1661.3		& \bidq		& 2.9 $\pm$ 0.7  	 \\
1729.6		& \bidq		& 9.3 $\pm$ 0.7  	 \\
1764.5		& \bidq		& 53.5 $\pm$ 1.7  	 \\
1847.4		& \bidq		& 6.5 $\pm$ 0.7  	 \\
2118.5		& \bidq		& 3.8 $\pm$ 0.5  	 \\
2204.2		& \bidq		& 17.0 $\pm$ 0.8  	 \\
2448.0		& \bidq		& 4.4 $\pm$ 0.5  	 \\
\noalign{\smallskip}\hline
\end{tabular}
\label{tab:u-lines}
\end{center}
\end{table}

The \ths activity is basically unchanged relative to \qino, especially at high energies. This suggests that most of the observed \ths contamination is located in the cryostat and lead shields (i.e. the two structures present in both experiments). There is also a contribution from close sources, most likely the copper \holder, since we do not observe a large attenuation of the low-energy \gm lines relative to the high-energy ones. Moreover, if the tower data are divided into floors of crystals, we do observe a floor by floor dependence that is not consistent with the one predicted by simulations. In Monte Carlo simulations this behavior can be reproduced by a point source close to the crystal that recorded the highest \gm counting rate. However, the floor by floor dependence is not seen in the \alph region, so the source cannot be directly facing the detectors, and is most likely located in the 50\,mK copper shield. A similar anomaly was present in \qino.

The two strongest \gm emitters of the \u chain are the \rndd daughters: \pbduq and \bidq. All of the observed \u peaks are attributed to these two isotopes. 
Compared to \qino, the average \u activity is a factor two to three lower in \qz, revealing a better cleaning of the detector components and/or a better \rndd control. 
Indeed, the intensity of the observed lines, particular those from \bidq, vary from dataset to dataset by as much as a factor of 3. Since \rndd is a gas, it emanates from contaminated materials in the laboratory environment and its air concentration changes depending on ventilation conditions. 
As described in Section~\ref{sec:MC}, the most likely place for \rndd to enter the experimental setup is in the gap between the OVC and the \pbext. Occasionally, the nitrogen flushing system inside the acrylic glass box malfunctions and the radon level increases in this volume. This additional varying component of the \u contamination is well modeled by an increase in the contamination of the \cryoext. 

\begin{table*}[!htb]
\begin{center}
\caption{\gm lines that appear in \qz \mspecone but do not belong to \ths and \u chains.}
\begin{tabular}{rrccrcr}

\hline
Energy 		& Isotope	& Origin		& Decay mode		& Half life		& Q value	 & Rate \\
(keV)	& & & & & (keV)	& (counts/kg/y) \\
\hline
122.1		&\cocs		& Cu cosmogenic activation	& EC	& 271.7\,d	& 836	 &  13.3 \pom 2.1\\
144.8		&\teudcm	& \sbudc\ \bmin decay 		& IT	& 57.4\,d	& 145	&  54.8 \pom 2.8 \\
427.9		&\sbudc		& Te cosmogenic activation	& \bmin	& 2.8\,y	& 767	& 6.9 \pom 1.8 \\
433.9		&$^{108m}$Ag& Pb cosmogenic activation	& EC 	& 438\,y	&1922	& 4.6 \pom 1.6\\
439.0 		&$^{202}$Tl	& $^{202}$Pb EC decay		& EC 	& 12.3\,d	&1363	& 6.3 \pom 1.7\\
600.6		&\sbudc		& Te cosmogenic activation	& \bmin	& 2.8\,y	& 767	& 6.8 \pom 1.4 \\
614.3		&$^{108m}$Ag& Pb cosmogenic activation	& EC 	& 438\,y	&1922	& 6.6 \pom 1.4\\
661.7		&\csuts		& fallout					& \bmin	& 30.2\,y	& 1176	&  9.8 \pom 1.9\\
723.3		&$^{108m}$Ag& Pb cosmogenic activation	& EC 	& 438\,y	&1922	& 6.1 \pom 1.3\\
803.1		&\podd		& \pbdd \bmin decay			& \alph & 138.4\,d 	& 5408	& 12.8 \pom 1.7\\
834.8		&\mncq		& Cu cosmogenic activation	& EC	&312.5\,d	& 1377	& 12.1 \pom 2.3\\
1063.7		&\bidzs		& fallout					& EC	& 31.6\,y	&2398	& 10.9 \pom 1.2\\
1173.2		&\cosz		& Cu cosmogenic activation 	& \bmin	& 5.3\,y	& 2824	 & 51.8 \pom 1.7\\
1332.5		&\cosz		& Cu cosmogenic activation 	& \bmin	& 5.3\,y	& 2824	 & 50.8 \pom 1.6\\
1460.7		&\kq		& environmental 			& EC+\bmin& 1.248 $\times$ 10$^9$\,y & 1505	 & 302.6 \pom 3.3\\
1770.2		&\bidzs		& fallout					& EC	& 31.6\,y	&2398	& 3.6 \pom 0.7\\

\noalign{\smallskip}\hline
\end{tabular}
\label{tab:gamma-lines}
\end{center}
\end{table*}

In addition, we observe a line at \unit[803]{keV} due to \podd not in equilibrium with the \u chain. This is fully explained by  the \pbdd ($\tau_{1/2}$=\unit[22.2]{y}) bulk contamination of the \pbext. Before it was installed \unit[25]{y} ago, this lead had a \pbdd activity of \unit[16]{Bq/kg}, which is now reduced by a factor $\sim$2. This same line is also present in \qino\ data. The activity reported in Table~\ref{tab:gamma-lines} for the \unit[803]{keV} peak includes a contribution from a \bidq line at \unit[806]{keV} from the \u chain, that cannot be disentangled. 

\kq is a unique long-lived isotope that decays via both $\beta^{-}$ and electron capture, with a negligible $\beta^{+}$ branching. The signature of this contaminant is the single \gm line at \unit[1.46]{MeV}, which is produced in $\sim$10\% of the decays. The single \gm line is not sufficient to determine the activity in each volume, however an asymmetry in the rate along the \qz\ tower suggests that in addition to bulk contaminants there is a \kq (extended) source at the bottom of the cryostat.

A number of isotopes produced by cosmogenic activation are also identified. The most critical of these contaminants is \cosz. The coincidence of its two \gm lines in a single crystal produces a peak in the \bbz ROI (at 2505.7~keV). \cosz is primarily the result of fast neutron interactions on copper~\cite{SusanaCosmogenic}, but it can also be produced in tellurium~\cite{BarbaraCosmogenic,EttoreCosmogenic}. 
Exploiting a coincident \gm analysis we  can set an upper limit of \unit[$3 \times 10^{-7}$]{Bq/kg} for the \cosz concentration in the crystals, i.e. well below the level needed to explain the observed rate. Therefore most of the \cosz contamination is located in the copper components.

The main copper components are the \holder, \cryoint, and \cryoext. 
Following~\cite{SusanaCosmogenic}, we derive upper limits on their \cosz activities based on the time they spent at sea-level and underground. The time spent underground before the start of data taking was a few months for the \holder, \unit[14]{y} for the \cryoint, and \unit[25]{y} for the \cryoext. 
Assuming the $^{60}$Co to be in saturation (i.e. the condition in which the production rate equals the decay rate) in both \cryoint and \cryoext just before the underground storage, its activity today would be about 180 $\mu$Bq/kg for \cryoint and 42 $\mu$Bq/kg for \cryoext. These are used as upper limits for the real contamination. We have only a very rough estimate for the cosmic ray exposure of the copper \holder (since this is made of many parts, each with its own history of exposure and underground storage), which results in an expected activity of $\sim$50 $\mu$Bq/kg.
 
As expected, we also observe \mncq and \cocs from the activation of copper~\cite{CopperProducts}. Due to their short half-lives, they can only be located in the \holder. The nuclear fallout product \csuts is also present as it was in \qino, therefore it is located in the cryostat copper structures (since it can't contaminate ancient Roman lead). Due to the low intensity of the peak \cryoint and \cryoext spectra are degenerate with each other, and we use the first to account for this contamination. The only expected contaminant due to the cosmogenic activation of the crystals themselves is \sbudc and its daughter \teudcm; in fact, \sbudc is the only long lived isotope produced in tellurium with high cross-section~\cite{BarbaraCosmogenic}. 

There is evidence for cosmogenic activation of the Roman lead in the \pbint. We observe three $\gamma$s emitted by \aguzom (see Table~\ref{tab:gamma-lines}), a silver isotope produced by the neutron activation of silver impurities in the Roman lead with $\tau_{1/2}=$\,\unit[438]{y}. We observe a peak at \unit[439]{keV} that is ascribed to $^{202}$Tl, a daughter of the long-lived cosmogenic activation product $^{202}$Pb ($\tau_{1/2}=$\unit[52.5 $\times$ 10$^3$]{y}). The presence of $^{202}$Tl contaminant is confirmed by a gamma spectroscopy measurement of a \pbint sample. 

Finally, we observe the two high-energy \gm peaks at \unit[1063.7]{keV} and \unit[1770.2]{keV} from the decay of \bidzs\ without the more intense line at \unit[569]{keV}, suggesting that this fallout product contaminates the \pbext, as confirmed by \mcuorez simulations.

%% file: alfa.tex
\begin{figure}[htb]
	\begin{center}
		\includegraphics[width=.49\textwidth]{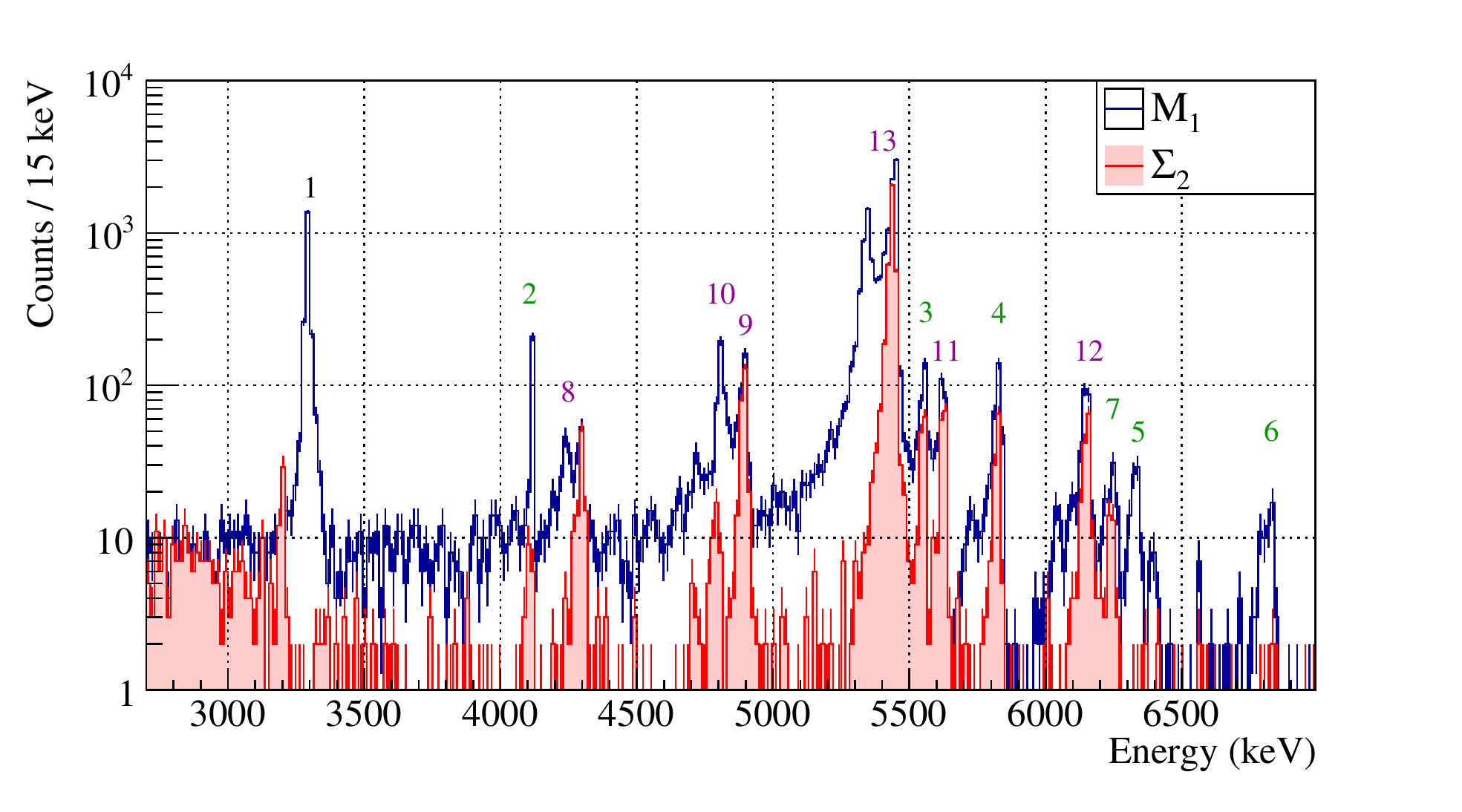}
		\caption{CUORE-0 \mspecone (blue) and \summspectwo (red) in the \alph region. The energy axis is calibrated on \gm. The peaks are labeled as follows: (1) $^{190}$Pt, (2) $^{232}$Th, (3) $^{228}$Th, (4) $^{224}$Ra, (5) $^{220}$Rn, (6) $^{216}$Po, (7) $^{212}$Bi, (8) $^{238}$U, (9) $^{234}$U and $^{226}$Ra, (10) $^{230}$Th, (11) $^{222}$Rn, (12) $^{218}$Po, (13) $^{210}$Po.}
		\label{fig:a_region}
	\end{center}
\end{figure}

The peaks observed in the \alph region of \qz (Fig.~\ref{fig:a_region}) come from the \u and \ths decay chains and \pt. Due to the short range of \alph particles and recoiling nuclei, the sources are contaminants close to or inside the bolometers: the \holder and the \crystal. The energy,  
multiplicity and intensity of the peaks can be used to efficiently constrain \u and \ths activities in these close components, simplifying the reconstruction of the more complicated \gm region. Furthermore they allow us to differentiate bulk and surface as well as \holder and \crystal contaminations.

Contaminants in the bulk of the \crystal produce \mspecone events with Gaussian peaks centered at the Q-value, since both the \alph and nuclear recoil are detected in the same crystal. 

\begin{figure}[htb]
	\begin{center}
		\includegraphics[width=.5\textwidth]{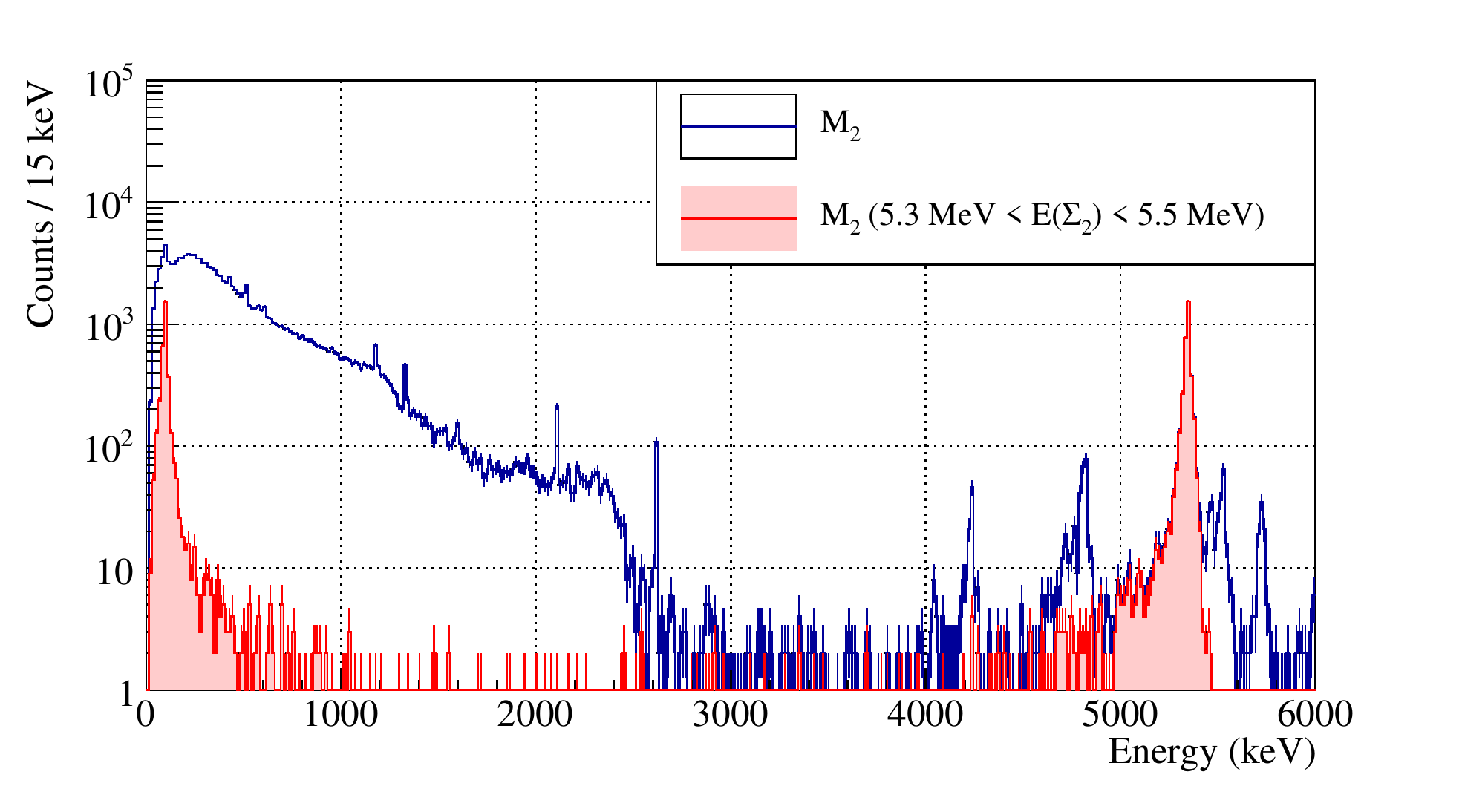}
		\caption{CUORE-0 \mspectwo spectrum (blue histogram, energy axis calibrated on \gm). By selecting \mspectwo events with a \summspectwo energy in the range \unit[5.3 - 5.5]{ MeV}, the peaks of $\alpha$s and nuclear recoils due to $^{210}$Pb on \crystal surfaces become visible (red histogram).}
		\label{fig:recoil}
	\end{center}
\end{figure}

\mspectwo events are a clear indication of the presence of contaminants on the surfaces of \crystal; they are produced when the \alph or the recoiling nucleus escapes the source crystal to enter one of the neighboring ones. These events reconstruct in the \summspectwo spectrum at the decay Q-value while in the \mspectwo spectra they produce two peaks each: one at the recoil energy ($\sim$\unit[70-100]{keV}) and one at the \alph energy ( $\sim$\unit[70-100]{keV} below Q-value); see Fig.~\ref{fig:recoil}. As the contaminants become deeper, the \alph loses more energy in the material where it originates from and a low-energy tail on the \alph peaks (or high-energy tail on the recoil peaks) becomes more pronounced. 

Contaminants on surfaces of \crystal also contribute to the \mspecone spectrum when both the \alph and recoiling nucleus are stopped within the source crystal or the escaping particles are not detected by another bolometer.

Contaminants in the \holder produce \mspecone events with at most the energy of the \alph. Again, depending on the depth, the line shape can vary from a peak with a strong low energy tail (shallow depth), to a flat continuum with no noticeable structures and extending far below the energy of the \alph (deep surface or bulk contamination). 

In general, the non-ideal behavior of the detector  complicates the reconstruction of \alph spectra. Some effects are well modeled in simulation: the energy threshold depletes \mspectwo in favor of \mspecone, and $\alpha$s are quenched with respect to $\gamma$s. 
Some effects are not fully understood, e.g. the broadening of all the peaks produced by surface contaminants, possibly explained by a different response of the bolometer to surface and bulk energy depositions. 
We choose a variable range binning to minimize these uncertainties as described in Section~\ref{sec:jags}.
\begin{table*}
\begin{center}
\caption{Activities of the most prominent $^{238}$U and $^{232}$Th alpha peaks for both surface and bulk contaminations. Line shape is not perfectly reconstructed, therefore activities are approximate. For the $^{210}$Po bulk contamination (which is decaying with $^{210}$Po half-life) we report the average activity recorded during CUORE-0 measurement. A few lines are depleted by pile-up  effects, therefore their activity cannot be directly measured. The energies listed in the second and third column are the nominal ones (no quenching included).}
	\begin{tabular}{c c c c c}
    	\hline
		Isotope & Q-value & $\alpha$-energy & Location & Activity \\
        \hline
        & [MeV] & [MeV] & Surface & [10$^{-2}$counts/cm$^{2}$/y] \\
		\hline
		\u & 4.30 & 4.20 & Crystal & 	1.5			\\ 
		$^{234}$U\,+\,$^{226}$Ra &$\sim 4.87$ & $\sim 4.78$ & Crystal &  3.6 \\ 
        \thdto & 4.77 & 4.69 & Crystal & 0.6			\\ 
        \rndd & 5.59 & 5.49 & Crystal & 2.0 \\ 
        $^{218}$Po & 6.12 & 6.00 & Crystal & 2.1  \\ 
        $^{214}$Po & 7.83 & 7.69 & Crystal & pile up: $\beta$($^{214}$Bi)+$\alpha$ \\
        \podd & 5.41 & 5.30 & Crystal+Holder & 86 \\ 
        \hline
        $^{228}$Th & 5.52 & 5.42 & Crystal & 1.8 \\ 
        \raddq & 5.79 & 5.69 & Crystal & 1.7 \\ 
        $^{220}$Rn\,+\,$^{216}$Po & -- & -- & Crystal &  pile up: $\alpha$+$\alpha$ \\
        \bidud & 6.21 & 6.05 & Crystal & 1.4 \\ 
        \podud & 8.95 & 8.79 & Crystal &  pile up: $\beta$($^{212}$Bi)+$\alpha$ \\
         \hline
         &  & & Bulk & [counts/kg/y]\\
        \hline
        \thdto & 4.77 & -- & Crystal & 10 \\ 
        \podd & 5.41 & -- & Crystal & $\langle 74 \rangle$ (time average) \\ 
        \hline
        \ths & 4.08 & -- & Crystal & 6.0 \\ 
		\hline
\end{tabular}
\label{tab:alfa}
\end{center}
\end{table*}

Table~\ref{tab:alfa} lists the isotopes producing the most prominent \alph peaks identified in CUORE-0 spectra. 

Surface contaminants in the \crystal are identified on the basis of the \summspectwo spectrum in Fig.~\ref{fig:a_region}. All the visible peaks belong to the \ths or \u chains. We observe breaks in secular equilibrium, particularly evident in the case of the \unit[5.4]{MeV} peak. This line, due to \podd ($\tau_{1/2}=138$\,d), is stable in time indicating an excess of \pbdd ($\tau_{1/2}=22.3$\,y).

The \mspecone peaks are for the most part due to the same surface contaminations of \crystal with four exceptions, discussed below.

The activity of the \unit[5.4]{MeV} line in the \mspecone spectrum decreases with a time scale consistent with the \podd half-life, indicating that there is a \podd bulk contamination of the \crystal. From this fit the average activity during CUORE-0 exposure is 2.36 $\mu$Bq/kg. This violation of the secular equilibrium of the \u chain is quite common in \teod~\cite{ccvr} due to the chemical similarities of tellurium and polonium.

\ths and $^{230}$Th (a \u chain product) produce sharp lines that can be explained by a bulk contamination of the \crystal . Most likely, these two long-lived isotopes are the only survivors of a \ths and \u contamination in the materials used for crystal production due to chemical affinity of thorium and oxygen.

Finally, we find evidence for a bulk \pt contamination of \crystal. The line in the CUORE-0 spectrum is shifted $\simeq$\unit[15]{keV} above the isotope Q-value of \unit[3249]{keV}, a much higher shift than the one observed for all the other lines in the \alph region. The crystals are grown in a platinum crucible. This contamination can be introduced during the crystal growth as a small grain of platinum locally modifying the detector response.

%% file: JAGS_pl.tex
The activities of the sources used for the background model are determined by fitting the observed \mspecone, \mspectwo, and \summspectwo spectra with a linear combination of simulated source spectra. 
The expectation value of the counts in the $i$-th bin of the experimental spectrum is given by:
\begin{equation}
\label{eq:jags}
\left\langle C_{i, \alpha}^{exp} \right\rangle = \sum_{j=1}^{57} N_j \left\langle C_{ij, \alpha}^{MC} \right\rangle 
\qquad {\alpha=\mathcal{M}_1, \mathcal{M}_2, \Sigma_2}
\end{equation}
where $\left\langle C_{ij, \alpha}^{MC} \right\rangle$ is the expectation value for the $i$-th bin of the simulated spectrum for the $j$-th source and N$_j$ is the unknown activity of the $j$-th source.  
To ensure sufficient statistics, both the experimental and simulated spectra are summed over all the active crystals. 

The fit is performed with a Bayesian approach using JAGS (Just Another Gibbs Sampler)~\cite{GibbsSampling,JAGS,BayesianSpectrum}. JAGS exploits Markov Chain Monte Carlo simulations to sample the \textit{joint posterior} probability distribution function (pdf) of the model parameters. 
Following Bayes' theorem the \textit{posterior} pdf is evaluated combining the \textit{likelihood} and the \textit{prior} distributions. The available data to define the \textit{likelihood}s are the observed counts in the bins of the experimental ($C_{i, \alpha}^{exp}$) and simulated ($C_{ij, \alpha}^{MC}$) spectra, both of which obey Poisson statistics.
Therefore, the \textit{joint posterior} pdf is:
\begin{equation}
\begin{aligned}
\begin{split}
Posterior(N_j, \left\langle C_{ij, \alpha}^{MC} \right\rangle  | C_{i, \alpha}^{exp}, C_{ij, \alpha}^{MC} )= \qquad \qquad \\
= \prod_{i, \alpha} Pois(C_{i, \alpha}^{exp}| \left\langle C_{i, \alpha}^{exp} \right\rangle) \times \prod_{j} Prior(N_j) \qquad \\ 
\times \prod_{ij,\alpha} Pois(C_{ij, \alpha}^{MC}| \left\langle C_{ij, \alpha}^{MC} \right\rangle) 
\times Prior(\left\langle C_{ij, \alpha}^{MC} \right\rangle)
\end{split}
\end{aligned}
\end{equation}

The \textit{prior}s for $N_j$, which describe our prior knowledge about source activities, are specified in Table~\ref{tab:SourceList}. In the case of a measured activity, we adopt a Gaussian \textit{prior} centered at the measured value with the measurement uncertainty as the width of the gaussian. For upper limits, we adopt a half gaussian with a width such that our 90\% upper limit is the 90\% value of the \textit{prior}. In all the other cases, we use a uniform non-informative \textit{prior} with an activity that ranges from 0 to an upper limit higher than the maximum activity compatible with the \qz data. Similarly, we use uniform \textit{prior}s over wide ranges for $\left\langle C_{ij, \alpha}^{MC} \right\rangle$. 

We chose a variable binning of the spectra to maximize the information content while minimizing the effects of statistical fluctuations and detector non-ideal behavior. Therefore, to avoid systematic uncertainties due to the lineshape, all the counts belonging to the same \gm or \alph peak are included in a single bin.
The minimum bin size in the continuum is \unit[15]{keV}, and bins with less than \unit[30]{counts} are merged with their immediate neighbor. The fit extends from \unit[118]{keV} to \unit[7]{MeV}. 
The threshold at \unit[118]{keV} is set to exclude the low-energy noise events (contaminating few datasets) and the nuclear recoil peak (that is mis-calibrated). In building the \summspectwo\ spectrum, we require that the energy of each event is above threshold. An exception is set for events with $E>2.7$~MeV in coincidence with events below the fit threshold, to correctly build-up the Q-value peaks in the \alph region of \summspectwo\ spectrum.

\input{Tabella.tex}

%% file: Tabella.tex
\begin{table*}
\begin{center}
\caption{List of the sources used to fit the \qz background data. 
The columns show (1) the name of the contaminated element, (2) the source index ($j$ in eq.~\ref{eq:jags}), and (3) the contaminant. If not otherwise specified, \ths, \u, and \pbdd refer to the whole decay chains in secular equilibrium, while the label ``only'' indicates that only the decay of the specified isotope is generated. For surface contaminants, the simulated depth is indicated in \ums. 
Column (4) reports the prior used in the fit, when not specified a non informative prior is used (see text for details). Column (5) reports the  posterior with the statistical error (limits are 90\% C.L.). Column (6) reports the range of systematic uncertainties 
(limits are 90\% C.L.). In the case of crystal sources, systematic uncertainties can arise from non-uniform contaminants in the different crystals.}	
\begin{tabular}{l|c|l|l|l|l}
\hline													
Component 	&	  Index  	&	Bulk sources	&	  Prior  [Bq/kg] 	&	  Posterior   [Bq/kg] 	&	
		Systematics [Bq/kg]	\\
\hline													
 \crystal	&	1	&	 $^{130}$Te -- \bbd 	&	   	&	3.43(9)\,$\times10^{-5}$	&	3.1\,$\times10^{-5}$	 --  	3.7\,$\times10^{-5}$	\\
 	&	2	&	 $^{210}$Po 	&	2.36(11)\,$\times10^{-6}$	&	2.39(11)\,$\times10^{-6}$			&		\\
 	&	3	&	 $^{210}$Pb 	&	   	&	1.37(19)\,$\times10^{-6}$	&	5.4\,$\times10^{-7}$	 -- 	2.2\,$\times10^{-6}$	\\
 	&	4	&	 \ths(only) 	&	   	&	7(3)\,$\times10^{-8}$	&	$<1.2\times10^{-7}$	\\
 	&	5	&	 $^{228}$Ra -- $^{208}$Pb 	&	   	&	$<3.5\times10^{-8}$	&	$< 	7.5\times10^{-8}$	\\
 	&	6	&	 \u\ -- $^{230}$Th 	&	   	&	$<7.5\times10^{-9}$	&	$<	3.6\times10^{-8}$	\\
 	&	7	&	 $^{230}$Th(only) 	&	   	&	2.8(3)\,$\times10^{-7}$	&				\\
 	&	8	&	 $^{226}$Ra -- $^{210}$Pb 	&	   	&	$<7.0\times10^{-9}$	&	$<	2.2\times10^{-8}$	\\
 	&	9	&	 $^{40}$K 	&	            	&	5.1(14)\,$\times10^{-6}$		&$<	8.2\times10^{-6}$	\\
 	&	10	&	  \cosz    	&	 $< 3.0\times10^{-7}$ 	&	$<5.1\times10^{-7}$	&				\\
 	&	11	&	 $^{125}$Sb 	&	   	&	9.6(4)\,$\times10^{-6}$	&	7.5\,$\times10^{-6}$	 -- 	 1.2\,$\times10^{-5}$	\\
 	&	12	&	 $^{190}$Pt 	&	   	&	2.00(5)\,$\times10^{-6}$	&	1.6\,$\times10^{-6}$	 -- 	 	2.3\,$\times10^{-6}$	\\
\holder 	&	13	&	 \ths 	&	 $< 2.0\times10^{-6}$ 	&	$<2.1\times10^{-6}$	&				\\
 	&	14	&	 \u 	&	 $< 6.5\times10^{-5}$ 	&	$<1.2\times10^{-5}$	&	$<2.2\times10^{-5}$	\\
 	&	15	&	 $^{40}$K 	&	7(2)\,$\times10^{-4}$	&	8(2)\,$\times10^{-4}$	&				\\
 	&	16	&	  \cosz    	&	5(1)\,$\times10^{-5}$	&	3.5(8)\,$\times10^{-5}$	&				\\
 	&	17	&	 $^{54}$Mn 	&	   	&	1.0(2)\,$\times10^{-5}$	&	$<	1.7\times10^{-5}$	\\
 	&	18	&	 $^{57}$Co 	&	   	&	2.9(3)\,$\times10^{-5}$	&	2.3\,$\times10^{-5}$	 -- 	 	3.7\,$\times10^{-5}$	\\
\cryoint  	&	19	&	 \ths 	&		&	$<1.5\times10^{-5}$	&	 $<3.5\times10^{-5}$	\\
 	&	20	&	 \u 	&		&	$<1.5\times10^{-5}$	&	 $<	3.9\times10^{-5}$	\\
 	&	21	&	 $^{40}$K 	&	   	&	1.1(3)\,$\times10^{-3}$	&				\\
 	&	22	&	  \cosz    	&	 $< 1.8\times10^{-4}$ 	&	2.4(10)\,$\times10^{-5}$			&		\\
 	&	23	&	 $^{137}$Cs 	&	   	&	9.9(15)\,$\times10^{-6}$	&				\\
\pbint  	&	24	&	 \ths 	&	 $< 4.5\times10^{-5}$	&	5.3(7)\,$\times10^{-5}$	&	1.7\,$\times10^{-5}$	 -- 	 6.6\,$\times10^{-5}$	\\
 	&	25	&	 \u 	&	 $< 4.6\times10^{-5}$ 	&	2.7(10)\,$\times10^{-5}$	&				\\
 	&	26	&	 $^{40}$K 	&	 $< 2.3\times10^{-5}$ 	&	$<2.4\times10^{-5}$	&	 $<	4.6\times10^{-4}$	\\
	&	27	&	 $^{108m}$Ag 	&		&	6.1(12)\,$\times10^{-6}$	&				\\
	&	28	&	 $^{202}$Pb	&		&	6(3)\,$\times10^{-6}$	&			\\
\cryoext  	&	29	&	 \ths 	&	   	&	$<1.2\times10^{-4}$	&	 $<	1.8\times10^{-4}$	\\
 	&	30	&	 \u 	&	   	&	2.4(6)\,$\times10^{-4}$	&	 $<	5.9\times10^{-4}$	\\
 	&	31	&	 $^{40}$K 	&	   	&	$<1.6\times10^{-3}$	&	 $< 2.6\times10^{-3}$	\\
 	&	32	&	  \cosz    	&	 $<4.2\times10^{-5}$ 	&	2.5(9)\,$\times10^{-5}$			&		\\
\pbext  	&	33	&	 \ths 	&	 $< 2.6\times10^{-4}$	&	3.1(3)\,$\times10^{-4}$	&	2.1\,$\times10^{-4}$	 --  3.5\,$\times10^{-4}$	\\
 	&	34	&	 \u 	&	 $< 7.0\times10^{-4}$ 	&	5.0(6)\,$\times10^{-4}$	&	3.5\,$\times10^{-4}$	 -- 	 6.2\,$\times10^{-4}$	\\
 	&	35	&	 $^{40}$K 	&	 $< 5.4\times10^{-3}$ 	&	3.1(5)\,$\times10^{-3}$	&	 			\\
 	&	36	&	 $^{207}$Bi 	&	   	&	5.9(5)\,$\times10^{-5}$	&	4.7\,$\times10^{-5}$	 -- 	 	7.2\,$\times10^{-5}$	\\
 	&	37	&	 $^{210}$Pb 	&	   	&	5.96(11)	&	5.4	 -- 	 	6.3	\\
\hline													
Component 	&		&	Surface sources	&	  Prior [Bq/cm$^2$]	&	  Posterior [Bq/cm$^2$]	&	
Systematics [Bq/cm$^2$]	\\
\hline													
 \crystal	&	38	&	 \ths(only)\,--\,0.01\,\ums 	&	   	&	3.0(10)\,$\times10^{-10}$	&				\\
 	&	39	&	 $^{228}$Ra -- $^{208}$Pb\,--\,0.01\,\ums 	&	   	&	2.32(12)\,$\times10^{-9}$	&	2.1\,$\times10^{-9}$	 -- 	 	2.7\,$\times10^{-9}$	\\
 	&	40	&	 \u\  -- $^{230}$Th\,--\,0.01\,\ums 	&	   	&	2.07(11)\,$\times10^{-9}$	&	1.8\,$\times10^{-9}$	 -- 	 2.2\,$\times10^{-9}$	\\
 	&	41	&	  $^{230}$Th  (only)\,--\,0.01\,\ums  	&	   	&	1.15(14)\,$\times10^{-9}$	&		 	\\
 	&	42	&	 $^{226}$Ra -- $^{210}$Pb\,--\,0.01\,\ums 	&	   	&	3.14(10)\,$\times10^{-9}$	&	2.9\,$\times10^{-9}$	 -- 	 3.5\,$\times10^{-9}$	\\
 	&	43	&	 $^{210}$Pb\,--\,0.001\,\ums 	&	   	&	6.02(8)\,$\times10^{-8}$	&	4.8\,$\times10^{-8}$	 -- 	 7.2\,$\times10^{-8}$	\\
 	&	44	&	 $^{210}$Pb\,--\,1\,\ums 	&	   	&	8.6(8)\,$\times10^{-9}$	&	7.2\,$\times10^{-9}$	 -- 	 1.1\,$\times10^{-8}$	\\
 	&	45	&	 $^{210}$Pb\,--\,10\,\ums 	&	   	&	$<2.7\times10^{-9}$	&	 $< 4.9\times10^{-9}$	\\
 	&	46	&	 \ths\,--\,10\,\ums 	&	   	&	7.8(14)\,$\times10^{-10}$	&				\\
 	&	47	&	 \u\,--\,10\,\ums 	&	   	&	$<3.3\times10^{-11}$	&	 $< 1.2\times10^{-10}$	\\
 \holder  	&	48	&	 $^{210}$Pb\,--\,0.01\,\ums 	&	&	2.9(4)\,$\times10^{-8}$	&	2.1\,$\times10^{-8}$	 -- 	 4.3\,$\times10^{-8}$	\\
 	&	49	&	 $^{210}$Pb\,--\,0.1\,\ums 	&	   	&	4.3(5)\,$\times10^{-8}$	&	3.1\,$\times10^{-8}$	 -- 	 5.1\,$\times10^{-8}$	\\
 	&	50	&	 $^{210}$Pb\,--\,10\,\ums 	&	   	&	$<1.9\times10^{-8}$		&$<		3.9\times10^{-8}$	\\
 	&	51	&	 \ths\,--\,10\,\ums 	&	   	&	5.0(17)\,$\times10^{-9}$		&$<	1.0\times10^{-8}$	\\
 	&	52	&	 \u\,--\,10\,\ums 	&	   	&	1.39(16)\,$\times10^{-8}$	&	8.4\,$\times10^{-9}$	 -- 	 	1.6\,$\times10^{-8}$	\\
\cryoint  	&	53	&	 $^{210}$Pb\,--\,0.01\,\ums 	&		&	1.4(7)\,$\times10^{-5}$		&$<	2.7\times10^{-5}$	\\
\pbint	&	54	&	 $^{210}$Pb\,--\,0.01\,\ums 	&		&	5.1(18)\,$\times10^{-5}$		&$<	8.2\times10^{-5}$	\\
\hline													
Component 	&		&	 Other  Sources  	&	  Prior [Bq]	&	  Posterior [Bq]	&	 	\\
\hline													
\emph{\unit[50]{mK} Spot}  	&	55	&	  \ths   	&	2.4(2)\,$\times10^{-4}$	&	2.41(18)\,$\times10^{-4}$				&	\\
\emph{Bottom Plate} \pbext	&	56	&	 $^{40}$K	&	16.8(2)	&	18(2)	&				\\
 Muons 	&	57	&		&	(see text)	&		&				\\
\hline																
\end{tabular}

\label{tab:SourceList} 
\end{center}
\end{table*}

%% file: systematics.tex
\begin{figure*}[htb!]
	\begin{center}
	\includegraphics[width=0.9\textwidth]{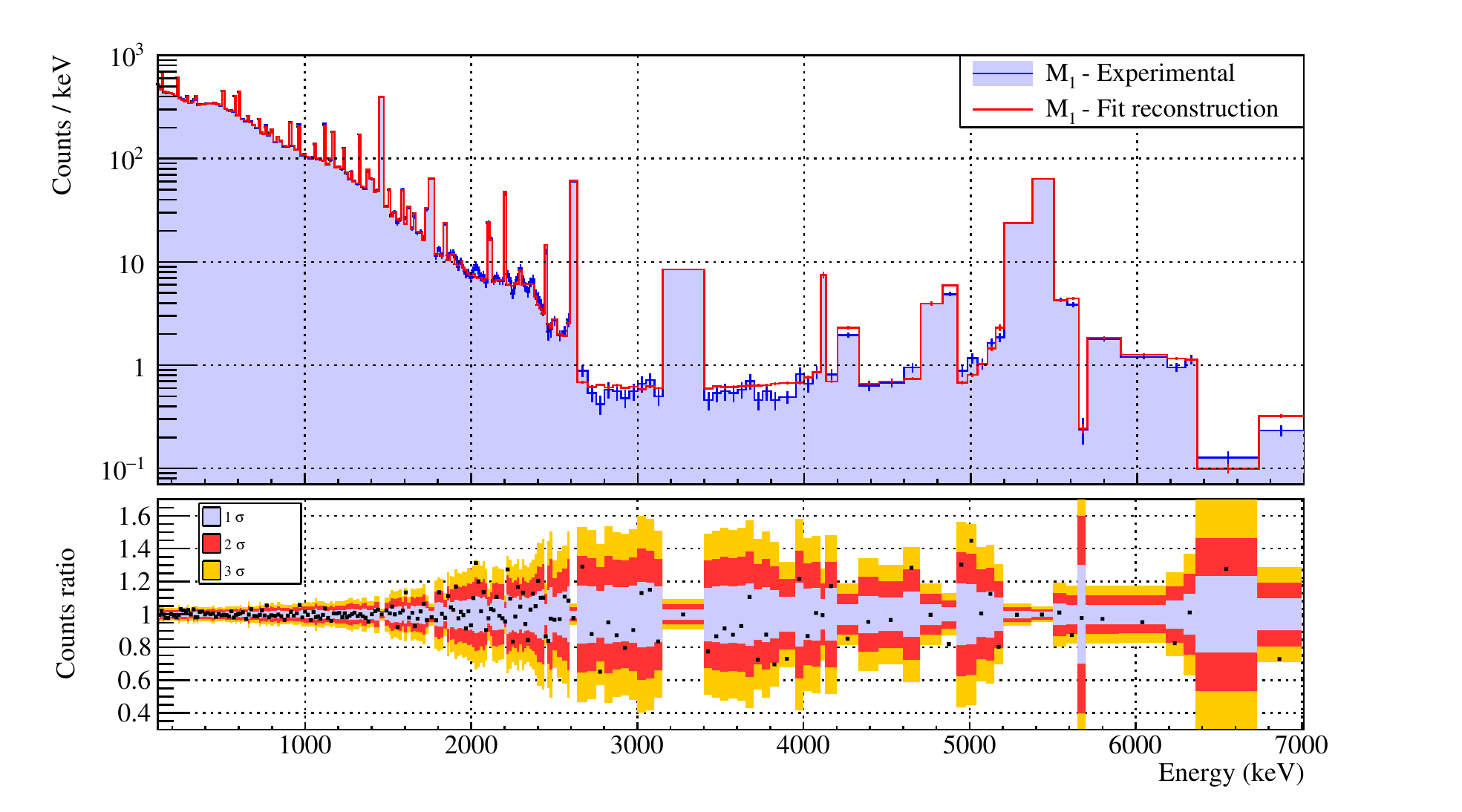}
	\end{center}
	\caption{Comparison between the experimental \mspecone and JAGS reconstruction (top panel). In the bottom panel the bin by bin ratios between counts in the experimental spectrum over counts in the reconstructed one are shown; the corresponding uncertainties at 1, 2, 3 $\sigma$ are shown as colored bands centered at 1. Fit residuals distribution is approximately gaussian 
with $\mu= (-0.03 \pm 0.09)$ and $\sigma=(1.1 \pm 0.1)$.}	
	\label{Fig:JAGS-M1} 
\end{figure*}		

\begin{figure*}[htb!]
	\begin{center}
	\includegraphics[width=0.9\textwidth]{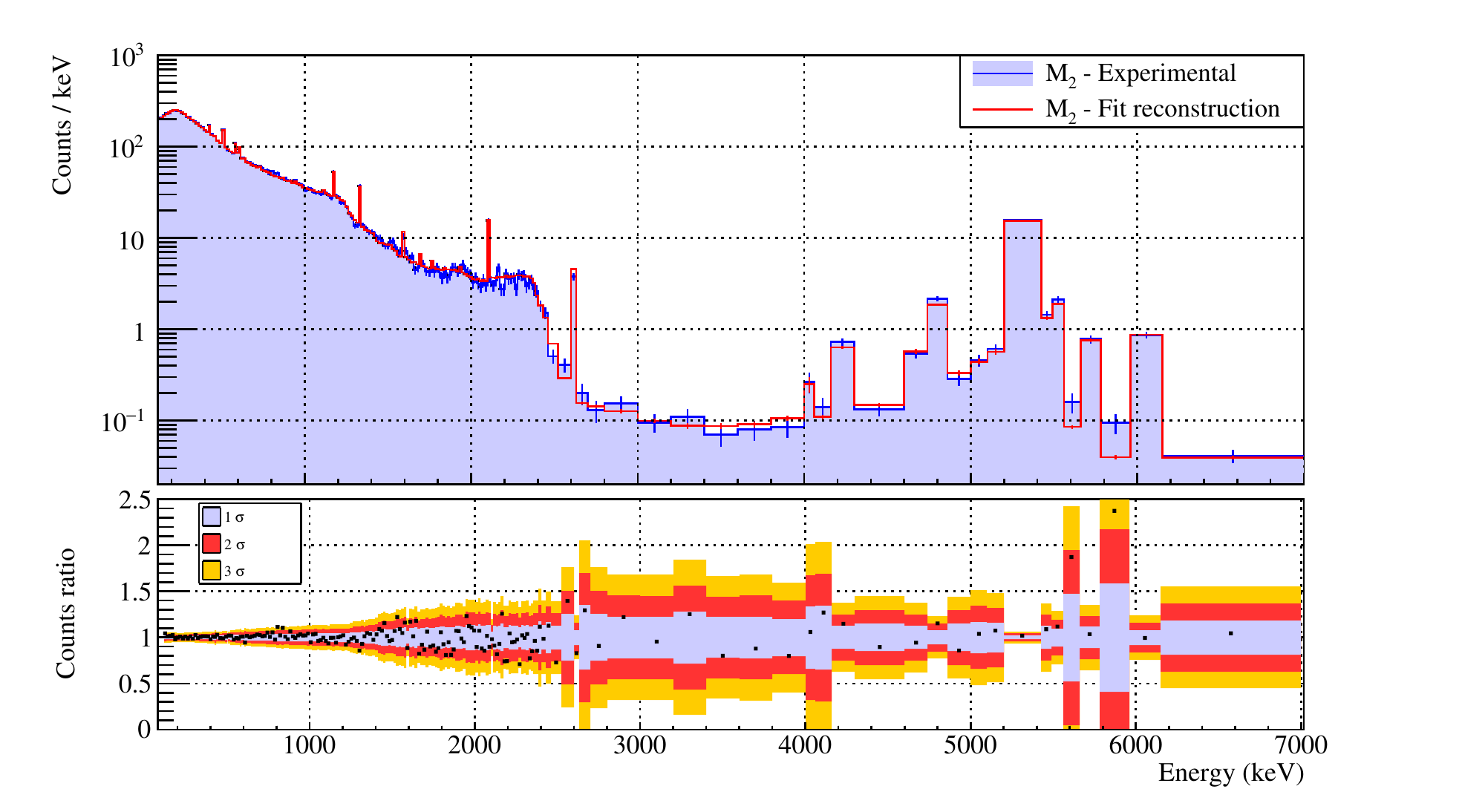}
	\end{center}
	\caption{Same as Fig.\,\ref{Fig:JAGS-M1} for \mspectwo. Fit residuals distribution is approximately gaussian 
with $\mu= (-0.13\pm0.08)$ and $\sigma=(1.00\pm0.08)$.}	
	\label{Fig:JAGS-M2} 
\end{figure*}		
	
\begin{figure*}[htb!]
	\begin{center}
	\includegraphics[width=0.9\textwidth]{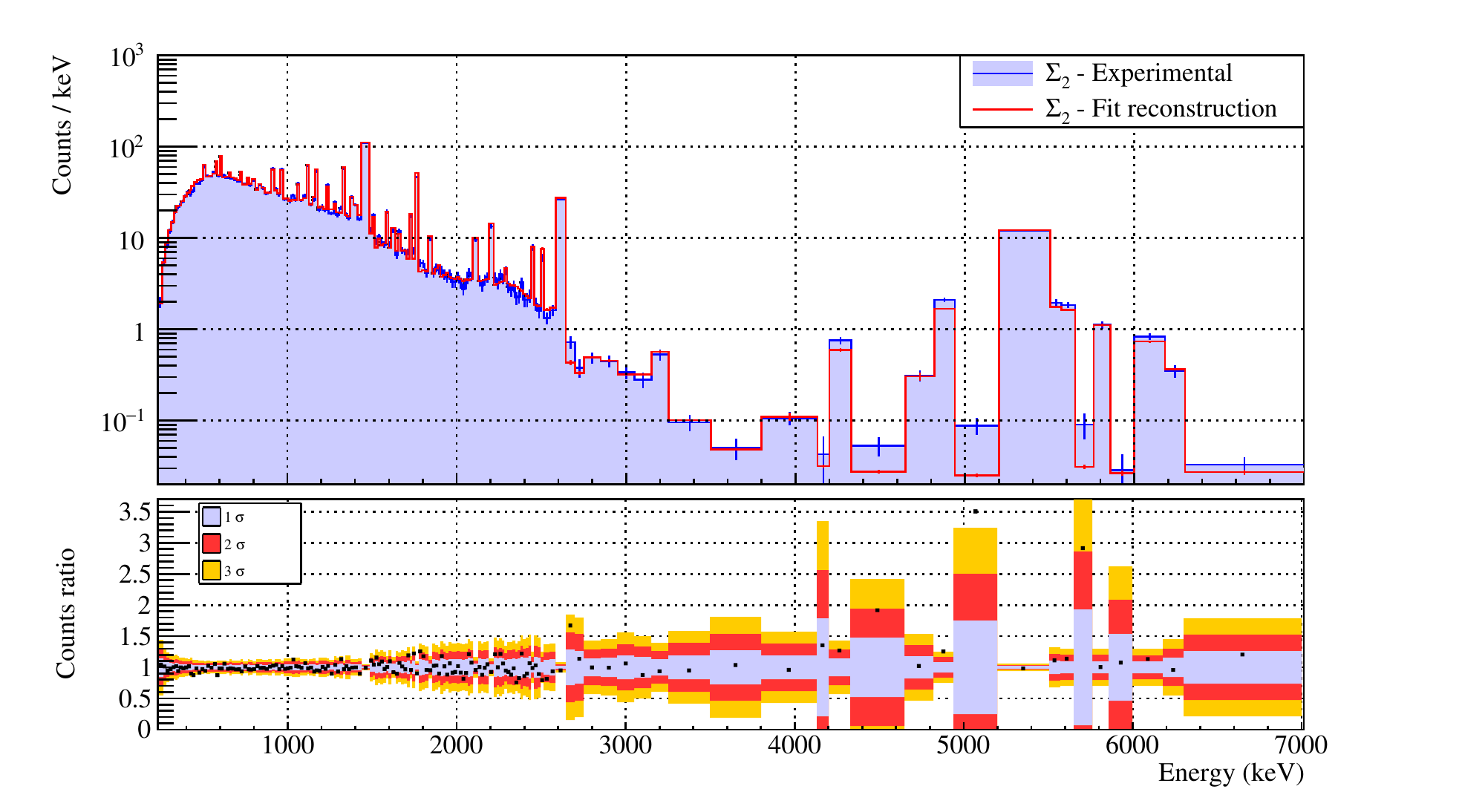}
	\end{center}
	\caption{Same as Fig.\ref{Fig:JAGS-M1} for $\Sigma_2$. Fit residuals distribution is approximately gaussian 
with $\mu$= (-0.09$\pm$0.09) and $\sigma$=(1.0$\pm$0.1).}	
	\label{Fig:JAGS-M2sum} 
\end{figure*}	

The \textit{reference} fit is the result of the fit to data from the total \unit[33.4]{kg$\cdot$yr} \teod exposure. The reconstructions of the experimental spectra are shown in Figs.~\ref{Fig:JAGS-M1},~\ref{Fig:JAGS-M2}, and~\ref{Fig:JAGS-M2sum} for the \mspecone, \mspectwo, and \summspectwo spectra, respectively. The fit results for the 57 free parameters are summarized in Table~\ref{tab:SourceList}. The marginalized posterior distributions are used to evaluate the central values and the statistical uncertainties of the activities, or to calculate 90\% upper limits for undetermined contaminations.

\begin{figure}[h!]
\begin{center}
\includegraphics[width=0.49\textwidth] {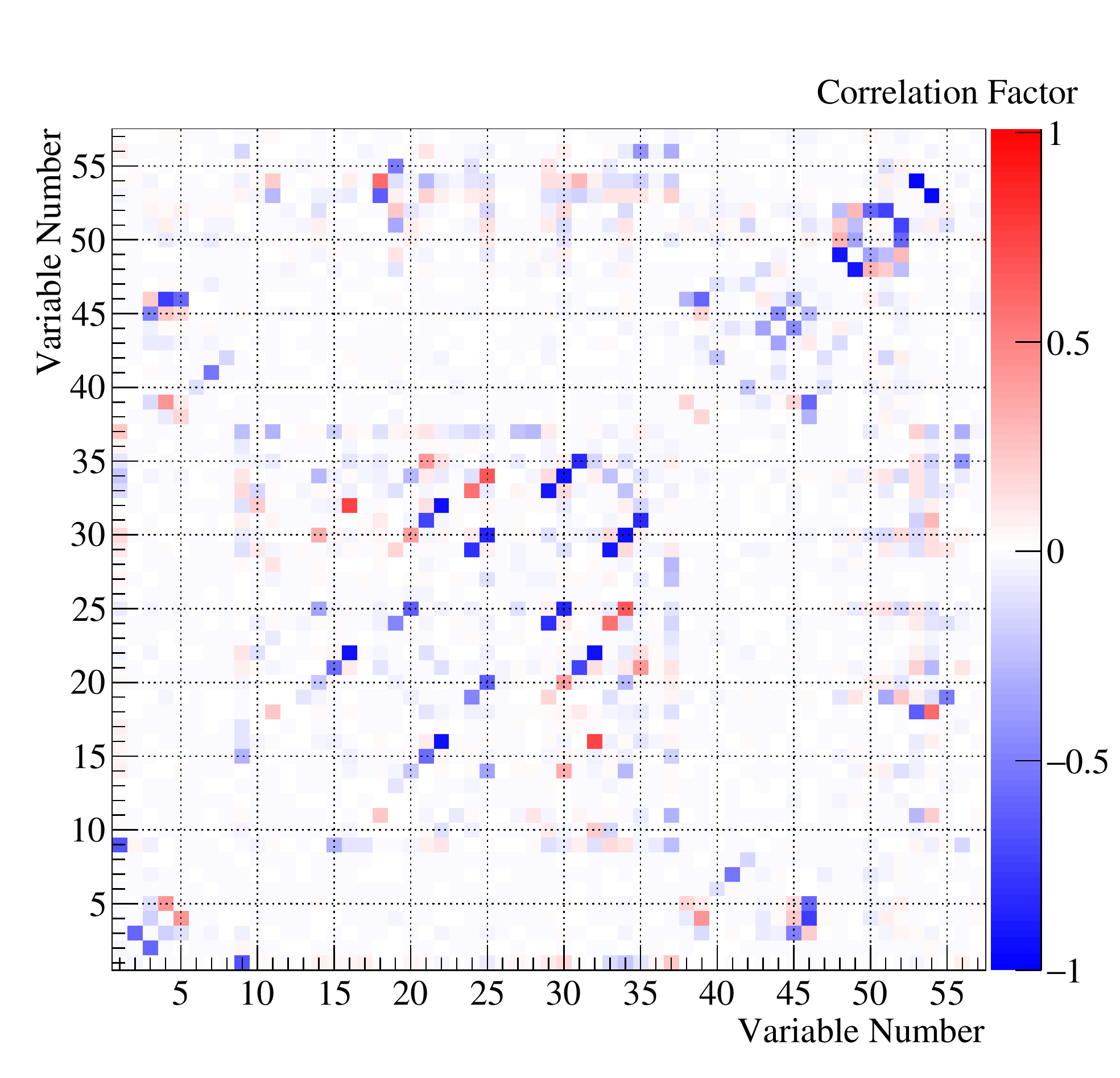}
\end{center}
\caption{Correlation matrix among the $N_j$ parameters. The list of sources is reported in Table~\ref{tab:SourceList}. Note the anti-correlation between the \bbd and the \kq activity in \textit{Crystal} bulk.}
\label{Fig:CorrelationMatrix}
\end{figure}

The correlation matrix of the 57 sources is illustrated in Fig.~\ref{Fig:CorrelationMatrix}. In general, most of the components used to fit the \alph region are not correlated to those used to reconstruct the \gm region.
As expected, the same contaminants in neighboring components of the experimental setup are highly correlated due to the similar spectra.

For the \mspecone, \mspectwo, and \summspectwo spectra, the normalized fit residuals have a Gaussian-like distribution with mean 0 and standard deviation 1. The reduced chi-square with 57 parameters and 478 degrees of freedom is 1.36. We do not expect perfect statistical agreement between the data and reconstruction since the uncertainties associated with the simulated spectra account for only the statistical fluctuation in the bin counts. They do not include the systematic uncertainties in the MC simulations. 

To check the stability of the background model, the dependence on priors, and the systematic uncertainties, especially those affecting the \bbd half-life, we run a number of different fits varying the binning, energy threshold, depth of surface contaminations, priors, list of background sources, and input data.

\begin{itemize}
\item \textbf{Binning.} We repeat the fit with different minimum bin sizes set to 5, 10, 20, and \unit[25]{keV}, and we test a uniform \unit[15]{keV} binning. The latter is the only case where the reconstruction is worse, because the line shape of \gm peaks is not perfectly modelled. The \bbd activity changes by less than 1\%. These tests also cover the systematics due to miscalibration \cite{Q0-analysis}.\\

\item \textbf{Fit Energy Threshold.} We run the fit with different energy thresholds ranging from 118 to \unit[500]{keV}, covering the region where the reconstruction of the calibration source is slightly worse (Sec.~\ref{sec:calibration}). The quality of fit reconstruction is unchanged and the \bbd activity variations are below 2\%.\\

\item \textbf{Contamination Depth Uncertainty.} We fit the \alph region with different depths to model surface contaminations. Several models perform similarly to the \textit{reference} fit, however the results of the background reconstruction, particularly the \bbd rate, are unaffected.\\

\item \textbf{Dependence on \textit{Prior} distributions.} To evaluate the systematic uncertainty related to the \textit{Prior} choice, we perform two JAGS fits. In the first fit, the half-Gaussian priors used in the case of upper limits on source activities are changed to uniform priors with the minimum at 0 and the maximum at $3\sigma$ above the upper limit. In the second fit, uniform non-informative priors are used for all components. In both cases, the global fit reconstruction is good and the \bbd result changes by $\sim$1\%.\\

\item \textbf{Selection of background sources.} In the reference fit there are 14 undetermined sources whose activity is quoted as upper limit. To check the fit stability against the removal of these sources, we run a \textit{minimum model} fit with only 43 sources. Once more, the global fit reconstruction and the \bbd result are not affected.\\

\item \textbf{Subset of Data.} We compare fit results obtained with various subsets of data. 

We search for time-related systematics by dividing the data into alternating datasets or grouping \emph{Rn-low} and \emph{Rn-high} datasets. Each study is performed with at least 1/3 of the total exposure. The \emph{Rn-low} and \emph{Rn-high} data are obtained by grouping the datasets in which the \bidq lines are more or less intense than the mean. This allows us to study if changes in the \bidq background influence the fit quality. The reconstruction results are compatible with the reference fit. The \u contamination in the \cryoext, which includes the air volume with the variable \rndd source, converges on results compatible with the different \rndd concentrations.

Finally, we investigate the dependence of the reconstruction on geometry by grouping the data by different floors: odd and even floors, upper and lower floors, the floors from 3 to 8 (central), and the complementary ones (peripherals). 
In this way, we explore the systematics due to model approximations. In Monte Carlo simulations we assumed contaminants to be uniformly distributed in each component of the experimental setup (except for the point sources) and we modelled the average performance of bolometers.
In all studies, the reconstruction is good, but we observe variations in the activities of the sources. In particular, the \bbd activity varies by about $\pm$ 10\%.\\

\end{itemize}

In the tests detailed above, the overall goodness of the ~fit remains stable, while we observe variations in the activities of the individual sources. These variations are used as an evaluation of the systematic uncertainty on the 57 source activities (Table~\ref{tab:SourceList}, sixth column).

There are caveats using the \textit{reference} fit results as an exact estimation of the material contamination. Indeed, degenerate source spectra allow us to use a single source to represent a group of possible sources. Examples are: the \holder that also accounts for the contribution of the \smallpart, surface contaminants in close components that are modeled with few representative depths, or bulk contamination in far components that also include surface ones.

%% file: results.tex
\section{\tect \bbd decay}
\label{sec:results}
\begin{figure}[h!]
\begin{center}
\includegraphics[width=0.49\textwidth]{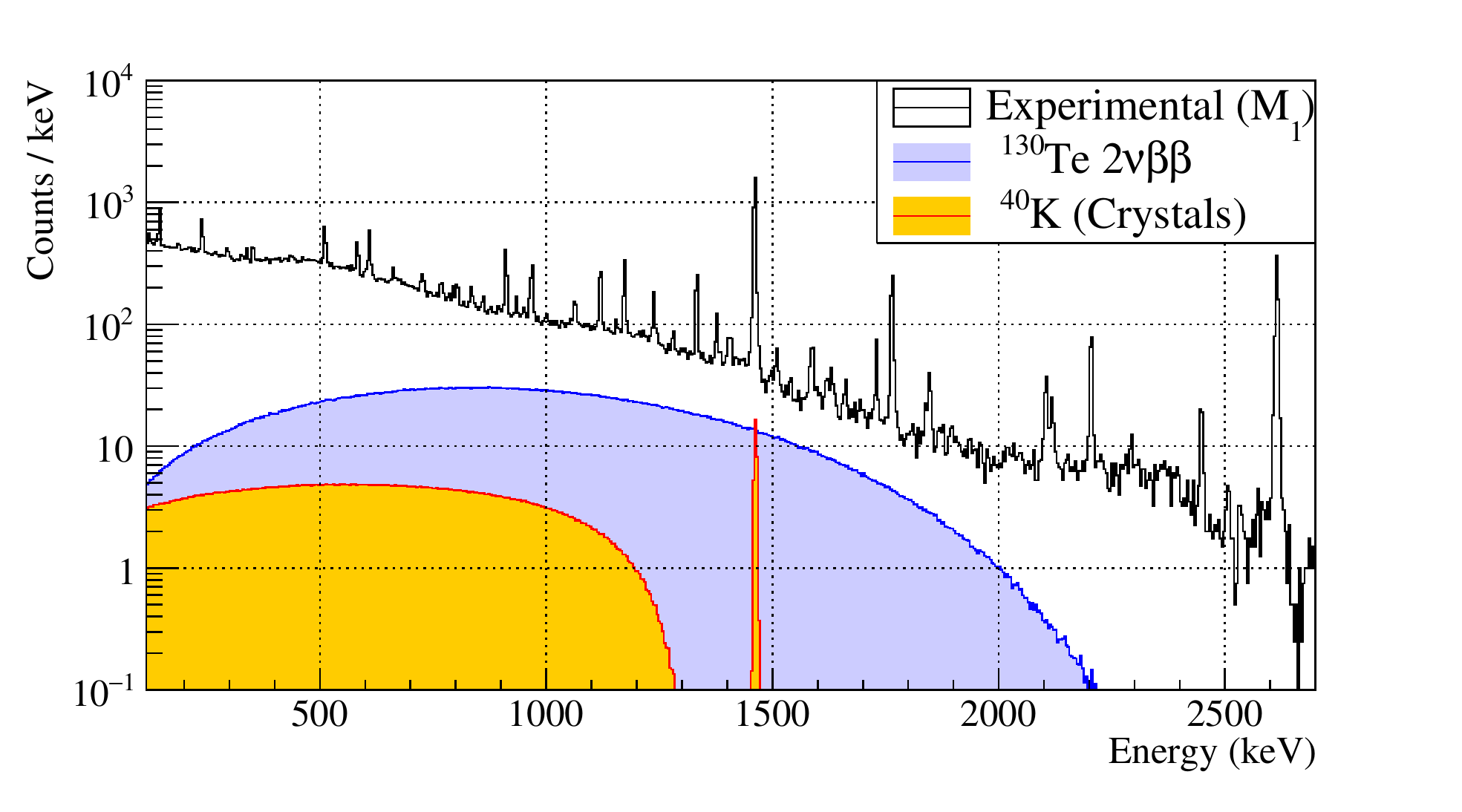}
\end{center}
\caption{CUORE-0 \mspecone compared to the \bbd contribution predicted by the \textit{reference} fit and the radioactive source that has the strongest correlation with \bbd, \kq in Crystal bulk.}	
\label{fig:2nu} 
\end{figure}

\begin{figure}[h!]
\begin{center}
\includegraphics[width=0.49\textwidth]{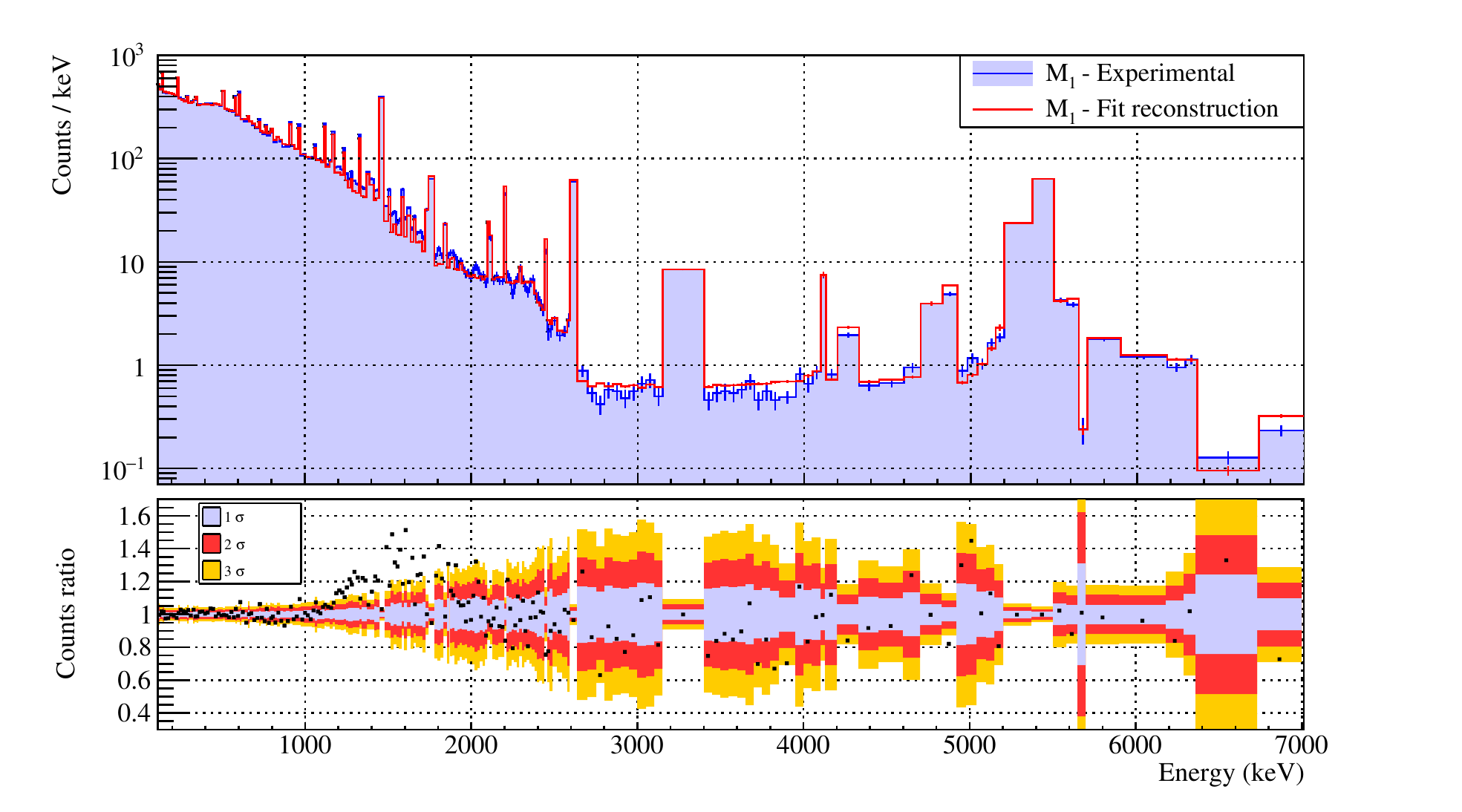}
\end{center}
\caption{CUORE-0 \mspecone compared to the reconstruction predicted by the fit without the \bbd source.}	
\label{fig:NO2nu} 
\end{figure}

The background reconstruction allows us to measure the \bbd of \tect with high accuracy. Fig.~\ref{fig:2nu} shows the fit result compared with the CUORE-0 \mspecone. \bbd produces $(3.27 \pm 0.08) \times 10^{4}$ counts, corresponding to $\sim$10\% of the events in the \mspecone \gm region from \unit[118]{keV} to \unit[2.7] {MeV}. As shown in Fig.~\ref{fig:NO2nu}, removing the \bbd component results in a dramatically poorer fit in this region.

\begin{figure}[h!]
\begin{center}
\includegraphics[width=0.49\textwidth]{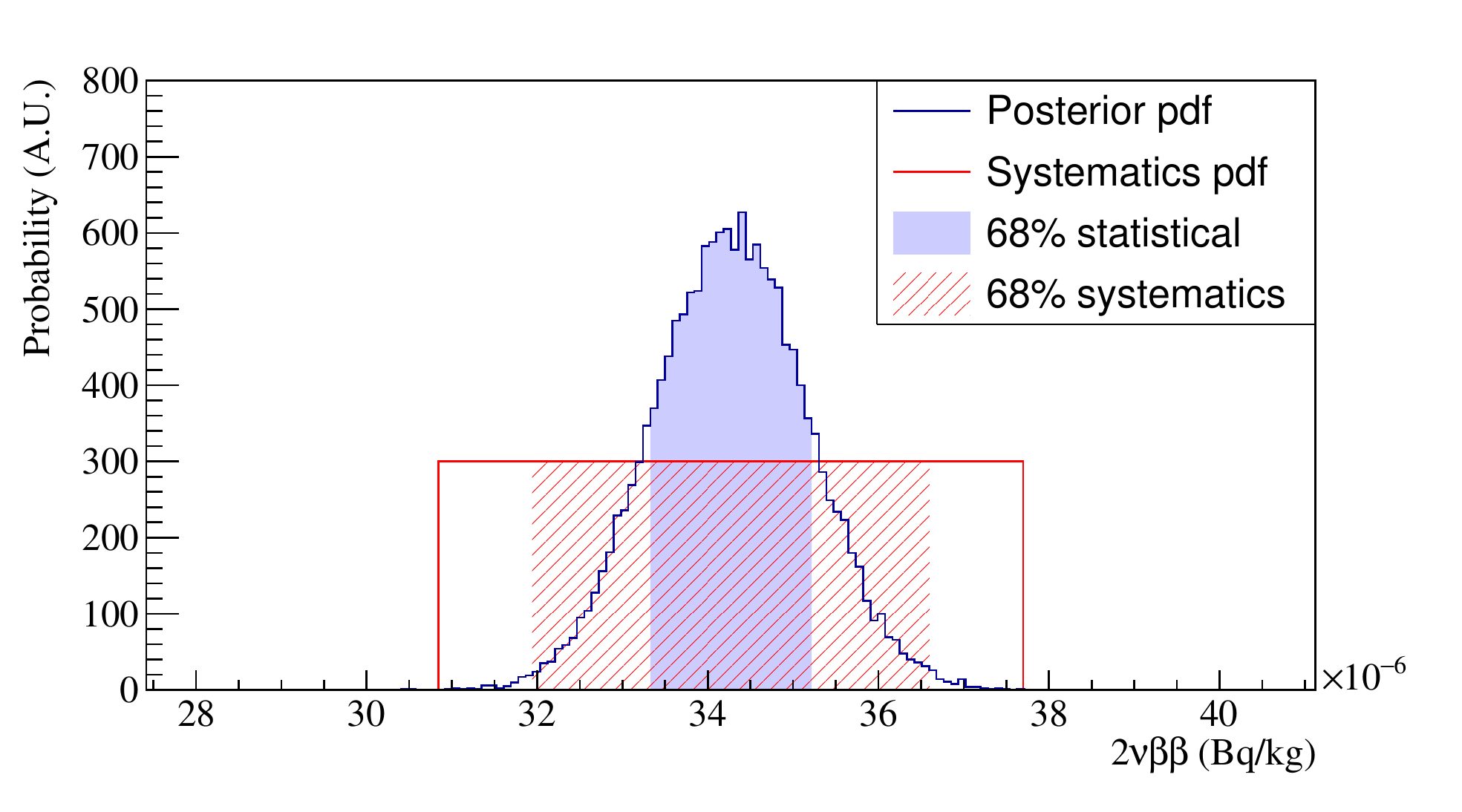}
\end{center}
\caption{\textit{Posterior} distribution of \bbd activity (blue) and systematic uncertainty range, represented as a flat distribution (red). The 68\% Confidence Intervals used to quote the statistical and systematic uncertainties are highlighted by colored areas.}	
\label{fig:2nuposterior} 
\end{figure}

The \bbd activity is (3.43 $\pm$ 0.09) $\times$ 10$^{-5}$ Bq/kg, with a statistical uncertainty that is amplified by the strong anti-correlation to the $^{40}$K contamination in crystal bulk (but not to other $^{40}$K sources). Indeed, this is the only case where the $\beta$ spectrum of $^{40}$K (having a shape that resembles that of \bbd) contributes to the detector counting rate. For all the other $^{40}$K sources, only the EC decay (branching ratio 89\%) contributes to the detector counting rate through the 1460~keV line and its Compton tail. 
The \textit{Posterior} for the \bbd activity as obtained from the reference fit is shown in Fig~\ref{fig:2nuposterior}. Also shown is the \textit{Posterior} associated to the fit bias. This is derived from systematic studies discussed in Sec.~\ref{sec:systematics} and is represented as a flat distribution. 
Fig~\ref{fig:2nuposterior} also shows the 68\% Confidence Intervals associated to statistical and systematic errors.

The half-life value obtained for \bbd is
\begin{center}
\Tdn~ = [8.2 $\pm$ 0.2 (stat.) $\pm$ 0.6  (syst.)] $\times$ 10$^{20}$~y 
\end{center}

\section{\tect \bbz region of interest}
\label{sec:ROI}
The signature of \tect \bbz  is a Gaussian line centered at \unit[2528]{keV}, the transition energy of the isotope, in \mspecone. Modeling the shape of the background in this region, especially possible subdominant peaks, and identifying the main sources of background is relevant not only for \qz, but also for the future evolution of \bbz searches with bolometers. 

\begin{figure}[h!]
\begin{center}
\includegraphics[width=0.49\textwidth]{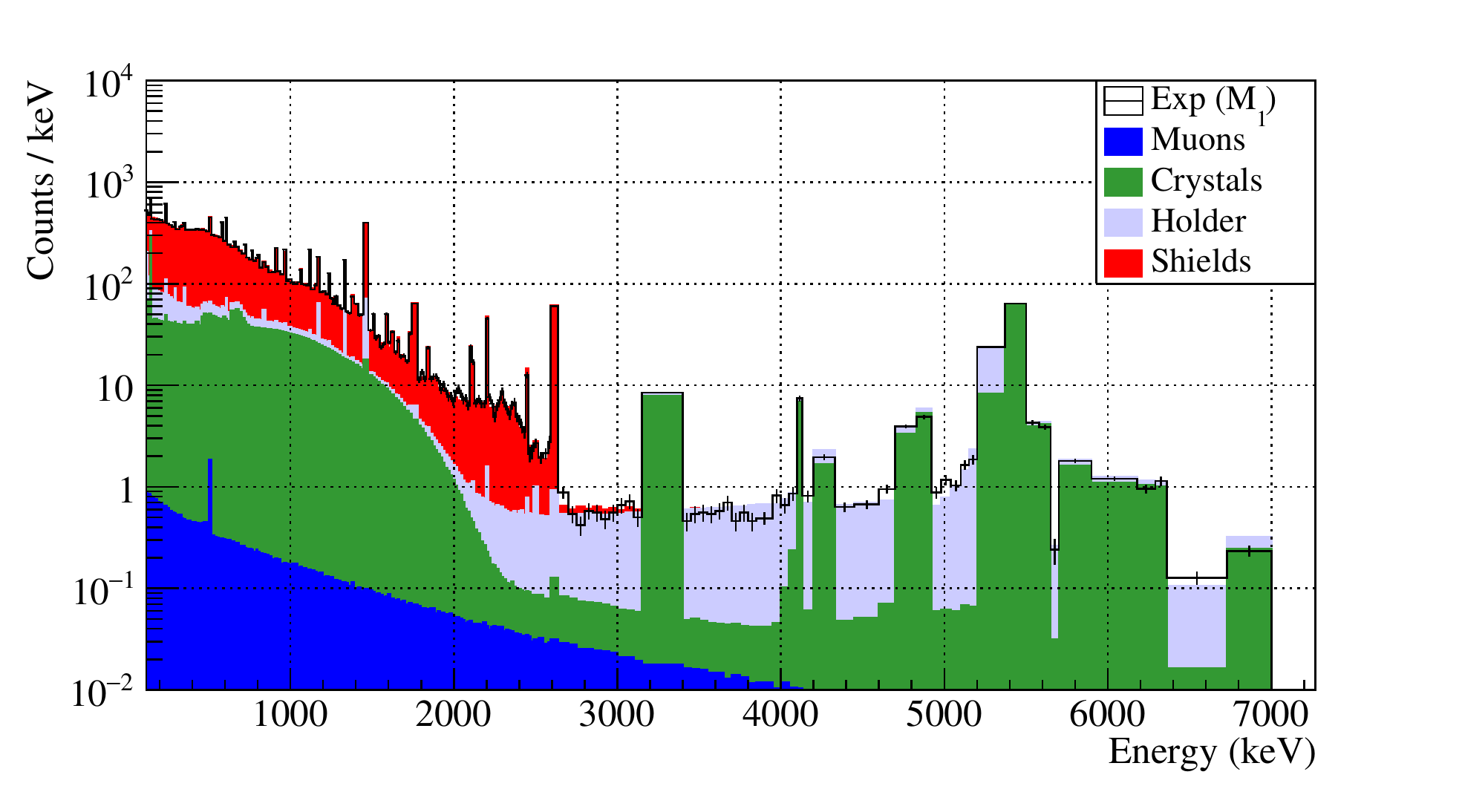}
\end{center}
\caption{Sources contributing to background reconstruction. ``Shields'' here stands for the sum of \cryoint, \cryoext, \pbint, and \pbext.}
\label{Fig:BkgCompM1} 
\end{figure}

It is useful to group the sources used for the fit into three major classes: the two elements that will be identical (though replicated 19 times) in CUORE, \holder and \crystal, and the element that will change, i.e. the cryogenic and radioactive shield systems (the sum of the \cryoint, \cryoext, \pbint, and \pbext). 
The contribution from these elements to the \bbz region of interest (\unit[2470-2570]{keV}) in \mspecone are shown in Fig.~\ref{Fig:BkgCompM1} and listed in Table~\ref{tab:ROIList}.
The largest contribution comes from the shields. This is mainly \ths contamination. The \holder is the second largest contributor due to degraded $\alpha$s from \u\allowbreak and \ths \textit{deep} surface contaminants. Bulk and shallow-depth contaminants account for less than 0.3\% of the background. A very small fraction of the background comes from \u, \ths, and \pbdd \crystal surface contaminants, and from muon interactions. The systematic uncertainties are negligible.

\begin{table}
\begin{center}
\begin{tabular}{l|c}
\hline															Component	&	Fraction [\%]			\\
\hline
Shields				&74.4 \pom1.3	 \\
Holder					&21.4 \pom0.7	 \\
Crystals				&2.64 \pom0.14	 \\
Muons					&1.51 \pom0.06	 \\
\hline															\end{tabular}
\caption{Sources contributing to the \bbz ROI. The flat counting rate in this region (i.e. excluding the \cosz sum peak) is $0.058\pm0.006$~counts/(keV~kg~y)~\cite{Us}. Column (2) reports the contribution of the different sources. ``Shields'' here stands for the sum of \cryoint, \cryoext, \pbint, and \pbext.}
\label{tab:ROIList} 
\end{center}
\end{table}

\begin{figure}[h!]
	\begin{center}
	\includegraphics[width=0.5\textwidth]{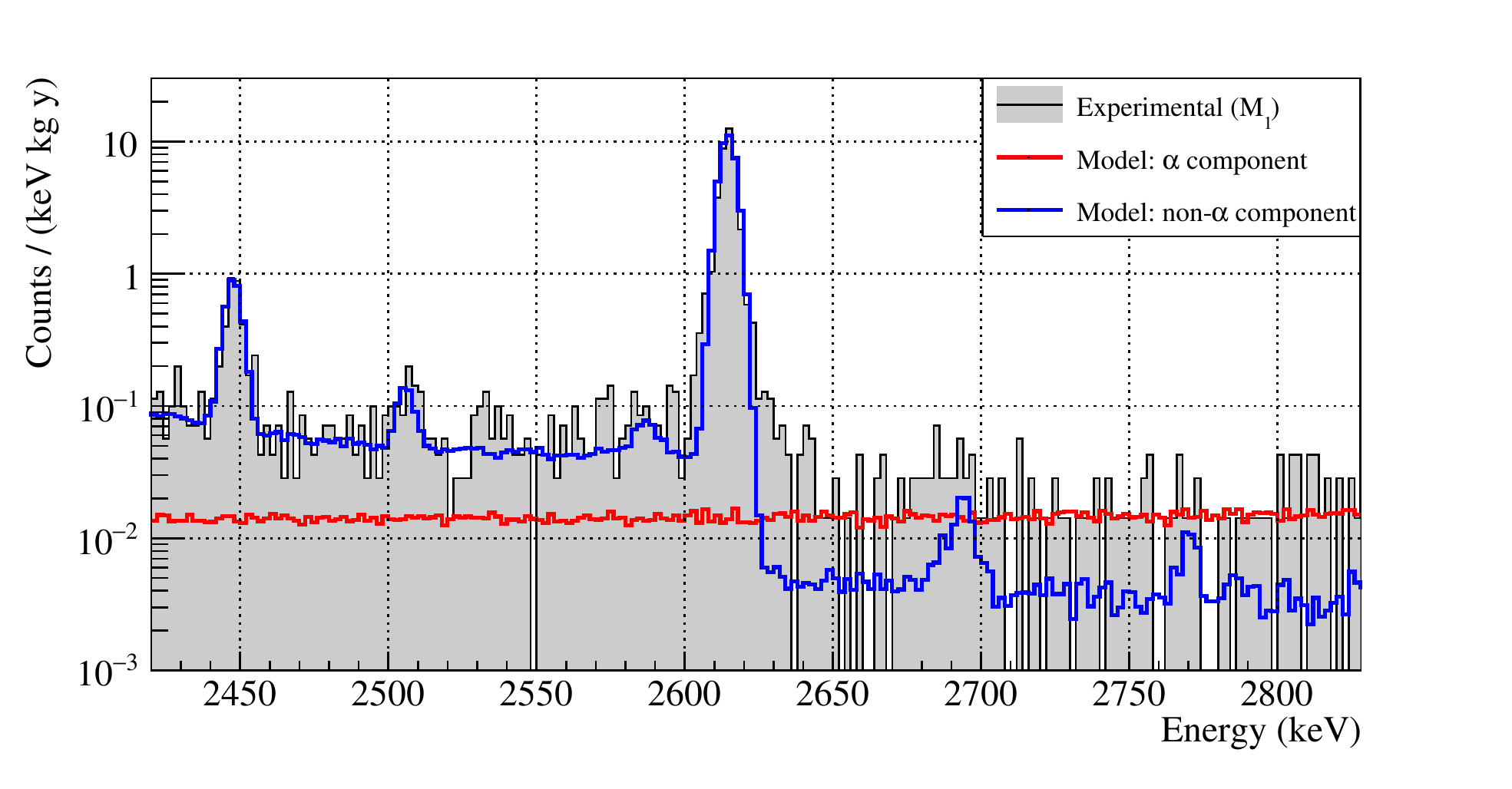}
	\end{center}
	\caption{Background reconstruction in the \bbz ROI. Events due to \alph particles (about 24\% of the ROI background) are shown in red. All the other events are shown in blue.}	
	\label{Fig:Q0alphaGamma} 
	\end{figure}

Finally, Fig.~\ref{Fig:Q0alphaGamma} shows a wider region centered around the ROI. This plot is produced by tagging the energy depositions where at least 90\% of the energy was deposited by \alph particles. We found $\sim24$\% of the ROI background was produced by \alph events. After reducing \gm backgrounds from the shields, these \alph events are expected to dominate the ROI rate in CUORE. This motivates the development of \alph particle discrimination for future bolometer-based experiments; see~\cite{CupidLOI} and references therein.
	
%% file: conclusion.tex
In this paper, we successfully reconstruct the \qz\allowbreak background using 57 sources modeled using a detailed Monte Carlo simulation. We find that 10\% of the \mspecone counting rate in the range [118$-$2700]~keV is unequivocally due to \tect \bbd decay. We measure its half-life to be:
\begin{center}
\Tdn~ = [8.2 $\pm$ 0.2 (stat.) $\pm$ 0.6  (syst.)] $\times$ 10$^{20}$~y. 
\end{center}
\noindent \sloppy Compared to previous results obtained from MiDBD [6.1 $\pm$ 1.4 (stat.) $^{+2.9}_{-3.5}$ (syst.)] $\times$ 10$^{20}$~y~\cite{MIDBD2n}, from \qino~\cite{Laura2n}, and from NEMO [7.0 $\pm$ 0.9 (stat.) $\pm$ 1.1 (syst.)] $\times$ 10$^{20}$~y~\cite{NEMO2n}, this is the most precise measurement to date. We find that the background rate in the \tect \bbz region of interest is dominated by the shields. 
This result gives us confidence that we are on track to achieve the requirements for CUORE.